\documentclass[
 amsmath,amssymb,
 aps,showkeys
]{revtex4-2}
\usepackage[normalem]{ulem}
\usepackage{graphicx}
\usepackage{dcolumn}
\usepackage{bm}
\usepackage{xcolor}
\usepackage{hyperref}

\definecolor{db}{HTML}{191586}
\definecolor{vi}{HTML}{5E1111}

\newcommand{\Ek}{\mathrm{Ek}}
\newcommand{\bs}[1]{\boldsymbol{#1}}

\begin{document}

\preprint{APS/123-QED}

\title{The stability of propagating plane inertial waves in rotating fluids}

\author{Valentin Skoutnev}
\thanks{valentinskoutnev@gmail.com}
\affiliation{Physics Department and Columbia Astrophysics Laboratory,  \\ Columbia University, 538 West 120th Street, New York, New York 10027, USA}%
\author{Aurélie Astoul}
\altaffiliation[Also at ]{IRAP, Université de Toulouse, CNRS, UPS, CNES, 14 Avenue Édouard Belin, F-31400 Toulouse, France.}
\affiliation{School of Mathematics, University of Leeds,\\ Leeds LS2 9JT, United Kingdom}%
\author{Adrian J. Barker}
\thanks{A.J.Barker@leeds.ac.uk} 
\affiliation{School of Mathematics, University of Leeds,\\ Leeds LS2 9JT, United Kingdom}%

\date{\today}

\begin{abstract}

Inertial waves transport energy and momentum in rotating fluids and are a major contributor to mixing and tidal dissipation in Earth's oceans, gaseous planets, and stellar interiors. However, their stability and breakdown mechanisms are not fully understood. We examine the linear stability and nonlinear breakdown of finite-amplitude propagating plane inertial waves using Floquet theory and direct numerical simulations. The Floquet analysis generalizes previous studies as it is valid for arbitrary perturbation wavelengths and primary wave amplitudes. We find that the wave vector orientation of the most unstable perturbations depends strongly on the wave frequency and weakly on the wave amplitude. The most unstable perturbations have wavelengths that are small relative to the primary wave wavelength for low wave amplitudes, but become comparable for large wave amplitudes. We then use direct numerical simulations to follow the nonlinear breakdown of the wave and examine how the wave energy is either dissipated in a forward cascade or accumulated into long-lived geostrophic modes. Simulations reveal that the conversion efficiency into geostrophic modes increases with increasing wave amplitude, as expected for pumping of geostrophic modes by nearly-resonant triadic interactions. We also find that the conversion efficiency increases with decreasing primary wave frequency, which may be due to the more efficient coupling of quasi-2D waves to geostrophic modes. These results on the stability and breakdown of single plane inertial waves provide an additional foundation for understanding the role of inertial waves in rotating turbulence, transport properties of inertial wave beams, and inertial wave propagation in more complex environments such as those with magnetic fields or shear flows.

\end{abstract}


\maketitle

\section{\label{sec:level1} Introduction \protect}

Rotating fluids support internal waves known as inertial waves (IWs), which result from the interaction between fluid inertia and the Coriolis force \citep{Greenspan1969}. In a fluid with a rotation rate $\Omega>0$, IWs are transverse, dispersive waves with frequencies $|\omega|\leq 2\Omega$ determined only by the angle of propagation relative to the rotation axis. They are fundamental in rotating flows because they mediate turbulent mixing and non-locally transport angular momentum and energy \citep{Staquet2002,davidson2012ten,Godeferd2015,alford2016near,joubaud2024internal}, thereby affecting the long-term dynamical evolution of rotating fluid bodies. 

The excitation, propagation, and dissipation of IWs have been studied in many natural systems, including stars, planets, and moons. Tidally excited IWs in star-planet systems can mediate angular momentum exchange as they propagate in convective zones and dissipate through wave-wave or wave-shear flow instabilities \citep{FBBO2014,AB2022,AB2023}. In planet-moon systems, dissipation of tidally-driven IWs can provide significant internal heating \citep{moons2019}, in addition to affecting their orbital and rotational evolution \citep[e.g.][]{Kerswell2002,OL2004,OL2007,BM2016,LC2020,B2020,B2022}. In terrestrial contexts, near-inertial waves in the upper oceans are a major contributor to the turbulent mixing that controls the depth of the oceanic mixed layer, impacting atmosphere-ocean coupling and the climate \citep{jochum2013impact}. Observational studies have successfully identified IWs in several natural systems, including the solar convective envelope \citep{Loptien2018,Gizon2021,Bekki2022,Hindman2022}, Earth's outer core \citep[e.g.][]{Aldridge1987}, and Earth's oceans \citep[e.g.][]{fu1981observations,hummels2020surface,zulberti2022mean}.  

Although IWs are important in geophysical and astrophysical applications, they have been less well studied than internal gravity waves (IGWs), which propagate in stably stratified fluids and also contribute to mixing in Earth's oceans \citep[e.g.][]{Staquet2002} and stellar interiors \citep[e.g.][]{BO2011}. These two wave types, however, derive from the same internal wave branch in a rotating and stratified incompressible fluid. As a result, they share many properties and, in particular, both are subject to triadic resonance instabilities (TRI) of small-amplitude waves, which transfer energy and momentum from a ``primary wave" to pairs of ``daughter waves" that satisfy temporal and spatial resonance conditions. Early studies of IGW stability focused on a subset of TRI known as parametric subharmonic instabilities (PSI) where the daughter wave frequencies are roughly half of the primary wave frequency \citep[for plane propagating IGWs, e.g.][]{Mied1976,Drazin1977,Klostermeyer1982,Sonmor1997}. One of the most comprehensive studies of IGW stability employed a numerical Floquet method applicable for any finite amplitude of the primary wave \citep{LR1996}. This approach also allows exploration of the stability of perturbations with wavelengths comparable to (or larger than) the primary wave, rather than being restricted to infinitesimally small wavelengths.

Similar efforts have been made to analyze the stability of IWs. Early work showed that standing IWs are generically unstable in both Cartesian \citep{Lifshitz1996,Miyazaki1998} and cylindrical geometries \citep{Kerswell1999}. IW attractors in 2D geometries have also been studied experimentally, numerically, and theoretically \citep{Bordes2012,Jouve2014}, where unstable perturbations were restricted to the plane of the rotation axis and the attractor. Relaxing this 2D assumption, \citet{Brunet2019} found that IW attractors are broadly unstable to 3D modes. For plane propagating IWs, the first analytical and experimental stability analysis  by \citet{Mora2021} found that the TRI for small-amplitude primary waves is also generally 3D---unstable modes efficiently redistribute energy into the horizontal plane normal to the rotation axis. Recently, \citet{AM2024} extended the linear stability analysis to finite-amplitude waves using a local stability approach appropriate for infinitesimally short-wavelength perturbations, characterizing the dependence of the most unstable perturbations on the primary wave amplitude and frequency. 

In this work, we extend the linear stability analysis of finite-amplitude, plane, propagating IWs to arbitrary wavelength perturbations and examine the nonlinear breakdown of IWs. We utilize the Floquet method of \citet{LR1996} for the stability analysis in Section~\ref{sec:FloqStabAnalysis} and supplement this with asymptotic analysis using the method of multiple timescales \citep[e.g.][ Appendix \ref{ap:PSI}]{Kerswell2002,OL2013,BO2014,CuiLatter2022}. We then use direct numerical simulations (DNS) of propagating IWs in Section~\ref{sec:DNS} to cross-validate the Floquet theory and examine the ultimate nonlinear breakdown of the wave, with a focus on the conversion efficiency of the wave energy into geostrophic modes.

\section{Floquet stability analysis of an inertial wave}
\label{sec:FloqStabAnalysis}
\subsection{Model and setup for the linear stability analysis} 
We introduce Cartesian coordinates $(x,y,z)$ and consider a homogeneous, incompressible fluid with density $\rho$ that is rotating at a constant rate $\bs{\Omega}=\Omega\hat{\bs{z}}$ around the unit vector $\hat{\bs{z}}$. The Navier-Stokes equation governing the fluid velocity field $\bs{u} (\bs{x} ,t )$ in the rotating coordinate system is
\begin{align}\label{eq:NS_1}
    \frac{\partial \bs {u} }{\partial t }+\bs {u} \cdot\bm\nabla \bs {u} +2\bs {\Omega}\times \bs {u} &=-\frac{1}{\rho}\bm\nabla  p +\nu{\bm\nabla }^2\bs {u},\quad \bs{u}=(u,v,w), 
\end{align}
where $\nu$ is the (constant) kinematic viscosity, and the pressure $p$ (which incorporates the centrifugal potential) is determined from the incompressibility condition $\bm\nabla \cdot\bs{u} =0$. Plane IWs are exact nonlinear solutions of Equation~(\ref{eq:NS_1}). In the inviscid limit, their velocity field is \citep{SW1999},
\begin{align}\label{eq:U_lab}
    \bs{U} =\mathrm{Re}[A\bs{h}_se^{\mathrm{i}(\bs{k}\cdot\bs{x} -\omega t ))}],\quad \bs{h}_s=\hat{\bm k}\times\hat{\bm\Psi}+\mathrm{i}s\hat{\bm\Psi},\quad \hat{\bm\Psi}=\frac{\hat{\bm k}\times\hat{\bm z} }{|\hat{\bm k}\times\hat{\bm z}|},\quad s=\pm1,
\end{align}
where $A$ is the amplitude of the wave and $\hat{\bm k}$ is the unit vector in the direction of the wavevector $\bs{k}$ (with magnitude $k=|\bs{k}|$). One can see that individual wave solutions have a vanishing nonlinear advection term, $\bs{U} \cdot \nabla \bs{U} =0$, and hence are exact solutions for any amplitude. The real frequency $\omega$ satisfies the dispersion relation
\begin{align}
    \omega=s 2\bs{\Omega}\cdot \hat{\bm k}=s2\Omega\cos\theta,\quad \textrm{with} \quad s=\pm1,
\end{align}
which depends only on the wave helicity $-s$ (defined based on the sign of $(\bm\nabla \times\bs{U}) \cdot \bs{U} $) and the angle $\theta$ between $\bs{\Omega}$ and $\hat{\bm k}$. For each wavevector $\bs{k}$, there are two waves with opposite helicities. Waves with $s=\pm1$ are right/left handed with a phase velocity along $\pm\hat{\bm k}$ and, in both cases, the velocity vector $\bs{U}$  is in the plane orthogonal to $\hat{\bm{k}}$
and at each point $\bs{x}$ rotates in time with rate $-2\Omega(\hat{\bm k}\cdot \hat{\bm z})$. 

The linear stability analysis of a plane IW is most convenient in a frame translating with the wave phase velocity. By rotating the coordinate frame to align $\hat{\bm z}$ with $\hat{\bs{k}}$ and applying a velocity boost along $\hat{\bm z}$ with the phase speed $\omega/k$, the base flow transforms from a periodic function of space and time coordinates in the rotating frame, $\bs{U}=\bs{U}(\bs{k}\cdot\bs{x} -\omega t)$, to a periodic function of a single spatial variable $\phi$ (the wave phase) in the comoving frame, $\bs{U}'=\bs{U}'(\phi)$.

The explicit transformation to the comoving frame is carried out as follows. Without loss of generality, consider a primary wave with a wavevector $\bs {k}=(k_x,0,k_z)$ in the $x $-$z $ plane, with $k_x\geq0$ and a real amplitude $A>0$. The units for space and time in the comoving coordinates are chosen based on the wavenumber and frequency of the primary wave.  The transformation from the rotating frame coordinates $(x ,y ,z )$ to the comoving dimensionless coordinates $(x',y',\phi)$ (with unit vectors $\hat{\bm{x'}}$, $\hat{\bm y'}$, and $\hat{\bm\phi}$) is then 
\begin{align}
    \label{eq:transformation_x}
    &x'=k_zx -k_xz ,\; y'=ky ,\; \phi=k_xx +k_zz -\omega t ,\; t'=|\omega| t , \;k=\sqrt{k_x^2+k_z^2},\\
    \label{eq:transformation_u}
    &u'=\frac{1}{|\omega|}\left(k_z u -k_x w \right),\; v'=\frac{k}{|\omega|}v,\; w'=\frac{1}{|\omega|}\left(k_x u +k_z w\right)-\frac{\omega}{|\omega|}, \;p'=\frac{k^2}{\rho\omega^2}p.
\end{align}
Transforming the Navier-Stokes equation to the comoving frame yields
\begin{equation}
    \label{eq:NS_2a}
    \frac{\partial \bs{u}'}{\partial t'}+\bs{ u}'\cdot\bm\nabla'\bs {u}'+\bs {\Omega}'\times \left(\bs{u}'+\frac{\omega}{|\omega|}\hat{\bm\phi}\right)=- \bm\nabla' p\,'+\mathrm{Ek}\bm\nabla^{'2}\bs{ u}',\quad \bm\nabla'=\left(\frac{\partial}{\partial x'},\frac{\partial}{\partial y'},\frac{\partial}{\partial \phi}\right),
\end{equation}
where $\Ek=\nu k^2/|\omega|$ is the Ekman number of the primary wave. A minor expense of using the comoving coordinates is a misalignment between the coordinate axes and the dimensionless rotation vector $\bs{\Omega}'=2\bs{\Omega}/|\omega|=(-k_x\hat{\bm x'}+k_z\hat{\bm\phi})/|k_z|$. However, the inviscid base flow $\bs{U}'$ of a plane IW is now simplified:
\begin{align}\label{eq:BaseFlow}
    U'&=A' \cos(\phi), \; V'=s A'\sin(\phi),\;W'=-\frac{\omega}{|\omega|},\; A'=\frac{Ak}{|\omega|}.
\end{align}
The new dimensionless amplitude $A'$ is essentially a Rossby number of the wave, quantifying the ratio of the advective rate $\sim A\,k$  to the wave frequency $|\omega|$. The horizontal components of the velocity ($U',V'$) trace out a right/left-handed corkscrew along $\hat\phi$ for $s=\pm1$, while the vertical component corresponds to uniform motion along the $\hat\phi$ axis.

We now consider the stability of perturbations $\bs{\widetilde u}'$ with small amplitudes $|\bs{\widetilde u}'|\ll |A'|$ on top of the inviscid base flow $\bs{U}'$. We neglect viscous damping of the primary wave because we are interested in the regime in which the growth rates of perturbations are considerably faster. This simplifying assumption also enables the use of Floquet theory for the linear stability analysis below. Substituting $\bs{u}'=\bs{U}'+\bs{\widetilde u}'$ and dropping quadratic terms $O(|\bs{\widetilde u}'|^2)$ gives the governing equations at first order for the perturbations,
\begin{align}
    \label{eq:NS_2b}
    \frac{\partial \bs{\widetilde u}'}{\partial t'}+\bs {U}'(\phi)\cdot\bm\nabla'\bs{\widetilde u}'+\bs{\widetilde u}'\cdot\bm\nabla'\bs {U}'(\phi)+\bs {\Omega}'\times \bs{\widetilde u}'&=-\bm\nabla' \widetilde{p}\,'+\mathrm{Ek}\bm\nabla^{'2}\bs{\widetilde u}'.
\end{align}

The partial differential equations in Equation (\ref{eq:NS_2b}) can be analyzed as follows. Note that their non-constant coefficients vary only with the $\phi$ variable. Hence, the horizontal $x'$ and $y'$ directions, and time $t'$, can be simplified by seeking solutions of the form
\begin{equation}\label{eq:Floquet_1}
    \widetilde{f}'=\mathrm{Re} \left[f^\dagger(\phi)e^{\mathrm{i}(\alpha x'+\beta y')+\sigma'  t'}\right],
\end{equation}
where $\widetilde{f}'$ is one of $\widetilde{u}'$, $\widetilde{v}'$, $\widetilde{w}'$, $\widetilde{p}\,'$, and the form of $f^\dagger$ is determined by Floquet theory. The horizontal wave vector components $\alpha$ and $\beta$ are real while the frequency $\sigma' $ is complex. Unstable modes are those with $\mathrm{Re}[\sigma' ]>0$. Substitution of Equation~\ref{eq:Floquet_1} reduces the system to a set of coupled ordinary differential equations in $\phi$:

\begin{align}
    (S-\Ek L_2)u^\dagger+w^\dagger \frac{\mathrm{d} U'}{\mathrm{d}\phi}-\frac{\cos\theta}{|\cos\theta|}v^\dagger&=-\mathrm{i}\alpha p^\dagger,\label{eq:NSx}\\
    (S-\Ek L_2)v^\dagger+w^\dagger \frac{\mathrm{d} V'}{\mathrm{d} \phi}+\frac{\cos\theta}{|\cos\theta|}u^\dagger+|\tan\theta|w^\dagger&=-\mathrm{i}\beta p^\dagger,\label{eq:NSy}\\
    (S-\Ek L_2)w^\dagger-|\tan\theta|v^\dagger&=-\frac{\mathrm{d} p^\dagger}{\mathrm{d} \phi},\label{eq:NSz}\\
    \mathrm{i}\alpha u^\dagger+\mathrm{i}\beta v^\dagger+\frac{\mathrm{d} w^\dagger}{\mathrm{d} \phi}&=0,\label{eq:cont} 
\end{align}
where $S=\sigma' +\mathrm{i}\alpha U'+\mathrm{i}\beta V'+W'\frac{\mathrm{d}}{\mathrm{d} \phi}$, $L_2=-\alpha^2-\beta^2+\frac{\mathrm{d}^2}{\mathrm{d}\phi^2}$, $0\leq\theta\leq\pi$, and the last equation is the incompressibility constraint. The coefficients $U'$ and $V'$ are $2\pi-$periodic, so Floquet theory implies that $f^\dagger$ takes the form 
\begin{equation}\label{eq:Floquet_2}
    f^\dagger(\phi)=e^{\mathrm{i}\gamma \phi}\hat{f}(\phi)=e^{\mathrm{i}\gamma \phi}\sum_{n=-\infty}^{n=\infty} f_n e^{\mathrm{i} n \phi},
\end{equation}
where $\hat{f}(\phi)$ is a $2\pi-$periodic function that is expanded in Fourier modes with complex coefficients $f_n$. Floquet theory allows $f^\dagger$ to be aperiodic since $\gamma$ need not take an integer value. However, $\gamma$ must be real for physical solutions that are bounded along $\phi$. Substitution of the Floquet ansatz (Equations \ref{eq:Floquet_1} and \ref{eq:Floquet_2}) and the base flow (Equation \ref{eq:BaseFlow}) into the governing equations for the perturbations (Equation \ref{eq:NS_2b}) reduces the problem to a set of coupled algebraic equations: 
\begin{align}
    \label{eq:FloquetSystem_1}
    (\sigma' -\Ek L_{2n}) u_n+\frac{\mathrm{i}\alpha A' }{2}(u_{n+1}+u_{n-1})-\frac{\beta s A' }{2 }(u_{n+1}-u_{n-1})-&\frac{\mathrm{i}s\cos\theta}{|\cos\theta|}(\gamma+n)u_n-\frac{\mathrm{i} A' }{2 }(w_{n+1}-w_{n-1})
    \nonumber\\
    &-\frac{\cos\theta}{|\cos\theta|}v_n =-\mathrm{i}\alpha p_n,\\\nonumber\\
    \label{eq:FloquetSystem_2}
    (\sigma' -\Ek L_{2n}) v_n+\frac{\mathrm{i}\alpha A' }{2 }(v_{n+1}+v_{n-1})-\frac{\beta sA' }{2}(v_{n+1}-v_{n-1})-&\frac{\mathrm{i}s\cos\theta}{|\cos\theta|}(\gamma+n)v_n +\frac{sA' }{2 }(w_{n+1}+w_{n-1})\nonumber\\
    &+\frac{\cos\theta}{|\cos\theta|}u_n+|\tan\theta|w_n=-\mathrm{i}\beta p_n,\\\nonumber\\
    \label{eq:FloquetSystem_3}
   (\sigma' -\Ek L_{2n}) w_n+\frac{\mathrm{i}\alpha A' }{2 }(w_{n+1}+w_{n-1})-\frac{\beta sA' }{2 }(w_{n+1}-w_{n-1})-&\frac{\mathrm{i}s\cos\theta}{|\cos\theta|}(\gamma+n)w_n-|\tan\theta|v_n=-\mathrm{i}(\gamma+n)p_n,\\ \nonumber\\
    \label{eq:incomp_2}
    \mathrm{i}\alpha u_n+\mathrm{i}\beta v_n+\mathrm{i}(\gamma+n) w_n&=0,
\end{align}
where $L_{2n}\equiv-\alpha^2-\beta^2-(\gamma+n)^2$. The linear stability of the wave can now be determined by solving this algebraic eigenvalue problem for the complex frequency $\sigma' $ as a function of the parameters of the primary wave $\{A',\theta,s\}$, the horizontal wave vectors of the perturbations in the comoving frame $\{\alpha,\;\beta\}$, and the Floquet parameter $\gamma$. For completeness, the energy conservation equation for this Floquet system is derived in Appendix~\ref{ap:EnergyConsrv}. Note that this equation set is similar to the analogous Floquet system for IGWs \citep{LR1996}, but with different linear restoring terms (buoyancy versus Coriolis) and a non-zero $\hat{y}$ component of the base flow for IWs which is not present for IGWs.

The eigenvalue problem given by Equations \ref{eq:FloquetSystem_1}--\ref{eq:incomp_2} has several symmetries that reduce the parameter space over which to search for unstable modes. Several symmetries are shared with the Floquet system for IGWs \citep{LR1996}, which has three symmetries: 1) symmetry for eigenvectors $\bs{V}_n\equiv [u_n,v_n,w_n,p_n]\to\bs{V}_n^*$ (where $*$ denotes the complex conjugate and $\bs{V}^*_n=\bs{V}_{-n}$), $(\alpha,\beta,\gamma)\to (-\alpha,-\beta,-\gamma)$, and $\sigma' \to {\sigma'}^*$, 2) invariance of $\sigma' $ to integer shifts in the Floquet parameter $\gamma\rightarrow \gamma+1$, and 3) invariance of $\sigma' $ in the $\beta$-plane, $\beta\rightarrow-\beta$ and $v_n\rightarrow-v_n$. The Floquet system for IWs satisfies the first two symmetries, but lacks the third because the non-zero $V'\neq0$ component of the base flow interacts with perturbations that have  $\beta\neq0$  (unlike for IGWs where $V'=0$). As a result, the Floquet system for IWs must be explored in the upper (or lower) half of the $\alpha-\beta$ plane, and for all $0\leq\gamma<1$.

\subsection{Floquet Solutions}
\label{sec:NumericalFloq_1}

\begin{figure}
\includegraphics[width=\linewidth]{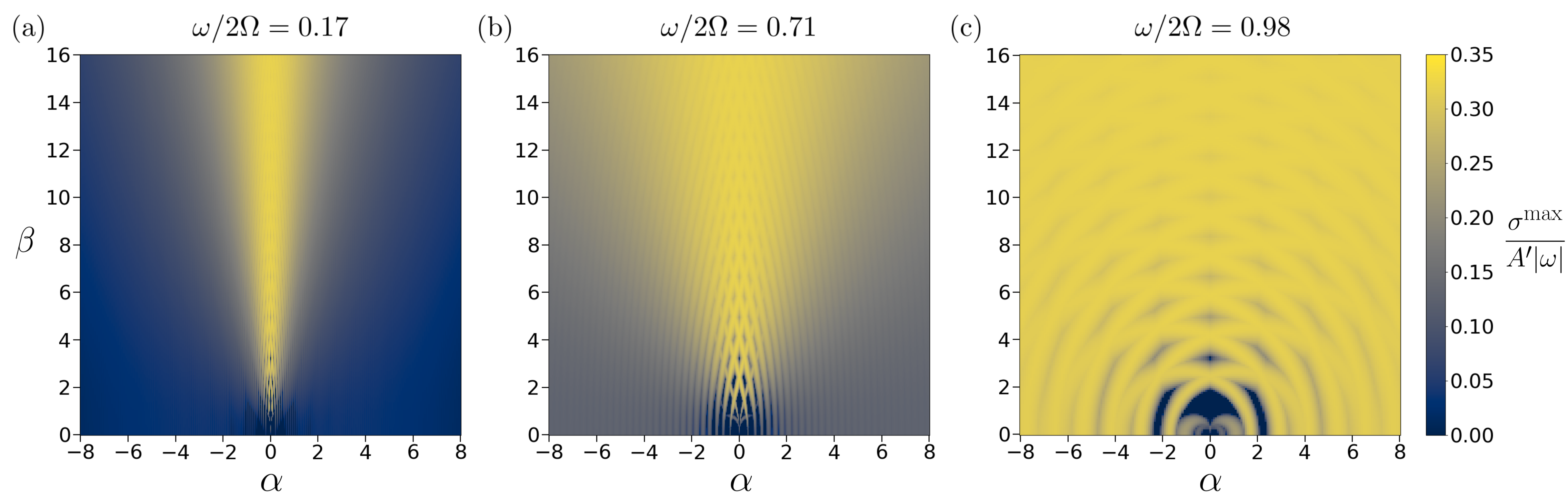}
\caption{\label{fig:omega_scan}
The maximum normalized growth rate $\sigma^{\max}/A'|\omega|$ of unstable perturbations predicted by Floquet theory as a function of the perturbation's horizontal wave vector components $\alpha,\beta$ for primary waves with frequencies $\omega/2\Omega=0.17,0.71,0.98$ (a-c), corresponding to angles $\theta=80^\circ,45^\circ, 10^\circ$. The Floquet parameter is fixed to $\gamma=0$ and all primary waves have the same normalized amplitude $A'=0.3$, helicity $s=1$, and $\Ek=10^{-6}$. }
\end{figure}

Solutions of the Floquet system in Equations (\ref{eq:FloquetSystem_1})-(\ref{eq:incomp_2}) are examined below using a numerical solver. The numerical solution to the infinite set of coupled equations is obtained by truncating the sum over $n$ with a finite number of modes $N$ in the range $-N/2< n\leq N/2$. The system is further reduced to $3N$ equations in the variables $u_n,v_n,w_n$ by eliminating the pressure $p_n$ using the incompressibility constraint in Equation~(\ref{eq:incomp_2}). The resulting system is a sparse $3N\times3N$ matrix whose eigenvalues and eigenvectors can be found using standard numerical software. The number of modes $N$ is chosen large enough that eigenvector components are negligible near the truncation mode $|n|\sim N/2$. The Ekman number $\Ek$  is chosen to be sufficiently small such that the viscous damping rate at all horizontal wave vectors is much smaller than the typical instability growth rate $\Ek(\alpha^2+\beta^2)\ll \sigma'\sim A'$. We focus on how the amplitude and frequency of the primary IW affect the growth rate and wave vector orientation of the most unstable perturbations.

\begin{figure}
\includegraphics[width=\linewidth]{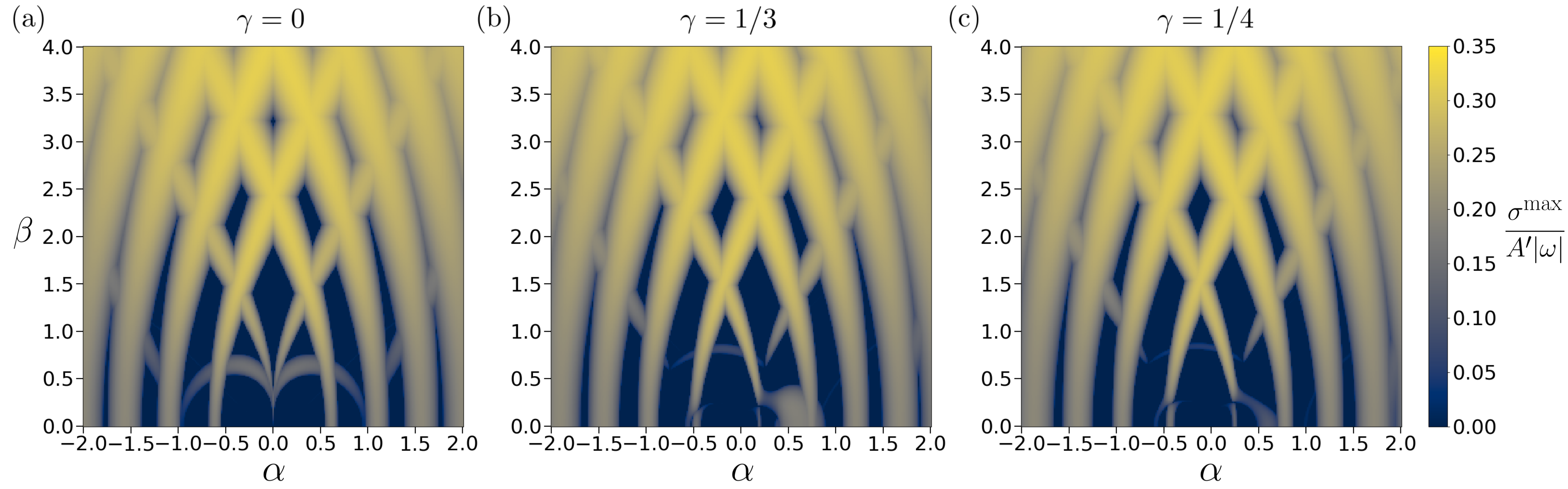}
\caption{\label{fig:gamma_scan} The maximum normalized growth rate of unstable perturbations with Floquet parameters $\gamma=0,1/3,1/4$ (a-c) for  the same primary wave with $A'=0.3$, $\omega/2\Omega=0.71$, $s=1$, and $\Ek=10^{-6}$. The panels focus on the origin where the effects of $\gamma$ are most pronounced since $\alpha\sim\beta\sim\gamma$. The $\gamma=0$ case in panel (a) is a zoomed-in region of panel (b) in Figure~\ref{fig:omega_scan}.
}
\end{figure}

We first vary the dimensionless wave frequency $\omega/2\Omega$, or equivalently the angle $\theta$ between $\bs{k}$ and $\bs{\Omega}$. Figure \ref{fig:omega_scan} shows Floquet stability diagrams for primary waves with frequencies $\omega/2\Omega=0.17,0.71,0.98$,  (angles $\theta=80^\circ,45^\circ, 10^\circ$), all with the same amplitude $A'=0.3$ and helicity $s=1$. The Floquet diagrams show the maximum normalized growth rate $\sigma^{\max}/A'|\omega|$ obtained at each horizontal wave vector with components $\alpha,\beta$ using the numerical Floquet solver with $N=128$ modes and a fixed Floquet parameter $\gamma=0$ (the dimensional growth rate is related to the non-dimensional growth rate by $\sigma=\sigma' |\omega|$). A characteristic feature of these plots is the ribbon patterns of the most unstable horizontal wave vectors. The center of each ribbon corresponds to an exact resonance of the PSI at small amplitudes $A'\ll1$, which is confirmed by an asymptotic analysis in Appendix \ref{ap:PSI} (see Figure~\ref{fig:PSI}). One can see that the orientations of the most unstable horizontal wave vectors have a strong dependence on the primary wave frequency. Lower frequency waves ($\bs k$ nearly orthogonal to $\bs\Omega$) are more unstable to perturbations with wavevectors in the $\hat{\bs{y}}$ direction (with $\alpha\sim0$), orthogonal to the $\bs k-\bs\Omega$ plane. In contrast, higher frequency waves ($\bs k$  nearly parallel to $\bs \Omega$) are unstable to perturbations whose wave vector orientation is more isotopically distributed in the plane perpendicular to $\bs k$. This is in agreement with the linear stability analysis in the short-wavelength limit in \cite[][see also Appendix~\ref{ap:PSI}]{Mora2021}. 

While the above results are for $\gamma=0$, a non-zero Floquet parameter does not change the trends observed with $\omega$ and $A'$. Physically, $\gamma$ accounts for modes that are not periodic along $\hat \phi$ with the primary wave's wavelength. Since the Floquet parameter is bounded as $0\leq\gamma<1$, varying $\gamma$ slightly shifts the instability ribbons in the $\alpha-\beta$ plane at large wavenumbers ($\alpha^2+\beta^2\gg\gamma^2$); however, it does not affect the maximum value of the growth rate (Appendix~\ref{ap:PSI}). This can be observed in Floquet diagrams (Figure~\ref{fig:gamma_scan}) where $\gamma$ is varied for the $\omega/2\Omega=0.71$ wave. 

We note a few additional properties of the Floquet diagrams that we have observed numerically. The diagrams all appear symmetric across the $\beta=0$ axis for $\gamma=0$, despite no obvious symmetry for $\alpha\rightarrow-\alpha$ in the Floquet system \ref{eq:FloquetSystem_1}--\ref{eq:incomp_2}. However, this symmetry is broken by non-zero values of the Floquet parameter $\gamma\neq0$. We also find that the most unstable perturbations always have $\alpha=0$ and that the local maxima in $\sigma^{\max}/A'|\omega|$ on the $\beta-$axis are independent of the wave frequency, which we confirmed with extra numerical scans. The main effect of increasing the wave frequency is thus the increase in the growth rates of perturbations with finite $\alpha$, resulting in more isotropic ribbon patterns. Lastly, flipping the primary wave helicity $s\rightarrow-s$ does not change the Floquet diagrams.

\begin{figure}
\includegraphics[width=\linewidth]{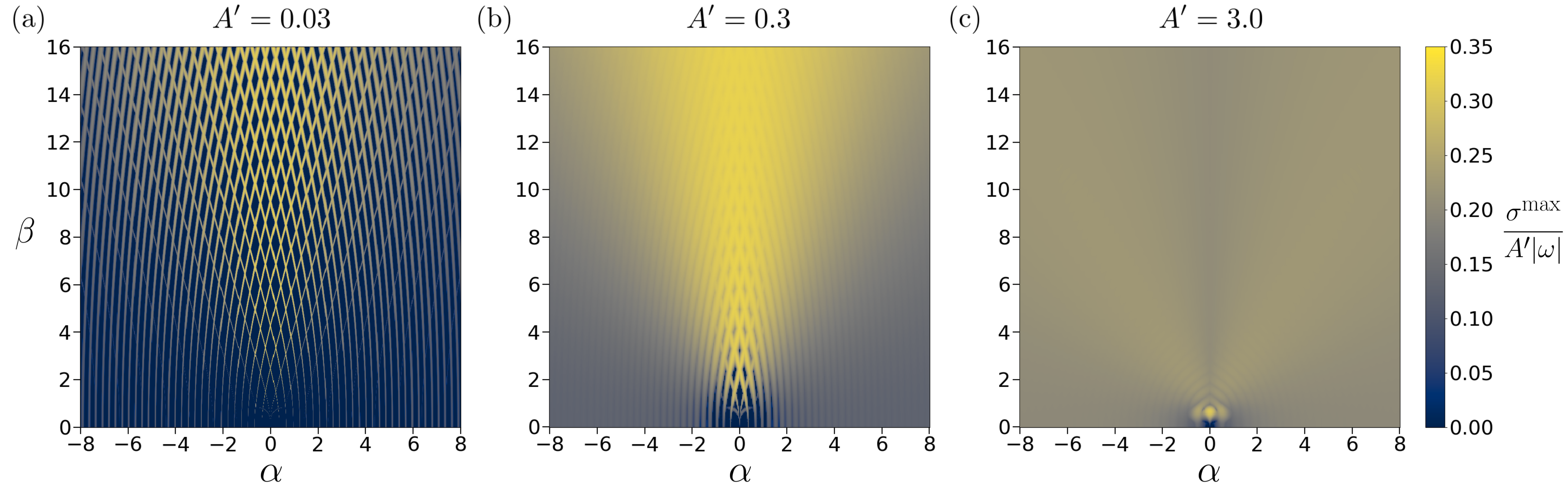}
\caption{\label{fig:A_scan}The maximum normalized growth rate of unstable perturbations for primary waves with amplitudes $A'=0.03,0.3,3.0$ (a-c),   all with the same frequency $\omega=\Omega$ (angle $\theta=60^\circ$), helicity $s=1$, and $\Ek=10^{-6}$.  The Floquet parameter is $\gamma=0$. 
}
\end{figure}

Next, we vary the primary wave amplitude $A'$. Figure \ref{fig:A_scan} shows Floquet stability diagrams for primary waves with amplitudes $A'=0.03,0.3,3.0$, all with the same frequency $\omega=\Omega$ (angle $\theta=60^\circ$) and helicity $s=1$. The main effect of moderately increasing the wave amplitude is the increased widths of the ribbons, as one can see by comparing panels (a) and (b) with $A'=0.03$ and $A'=0.3$, respectively. However, the predictions of the PSI break down at large amplitudes $A'\sim1$. This is evident in the case with $A'=3.0$ in panel (c), where the observed patterns no longer match the ribbons seen at low $A'$ and the fastest growing perturbations have horizontal wave vectors $\alpha=0$, $\beta\approx 1$ that are comparable to those of the primary wave.

\begin{figure}
\includegraphics[width=0.4\linewidth]{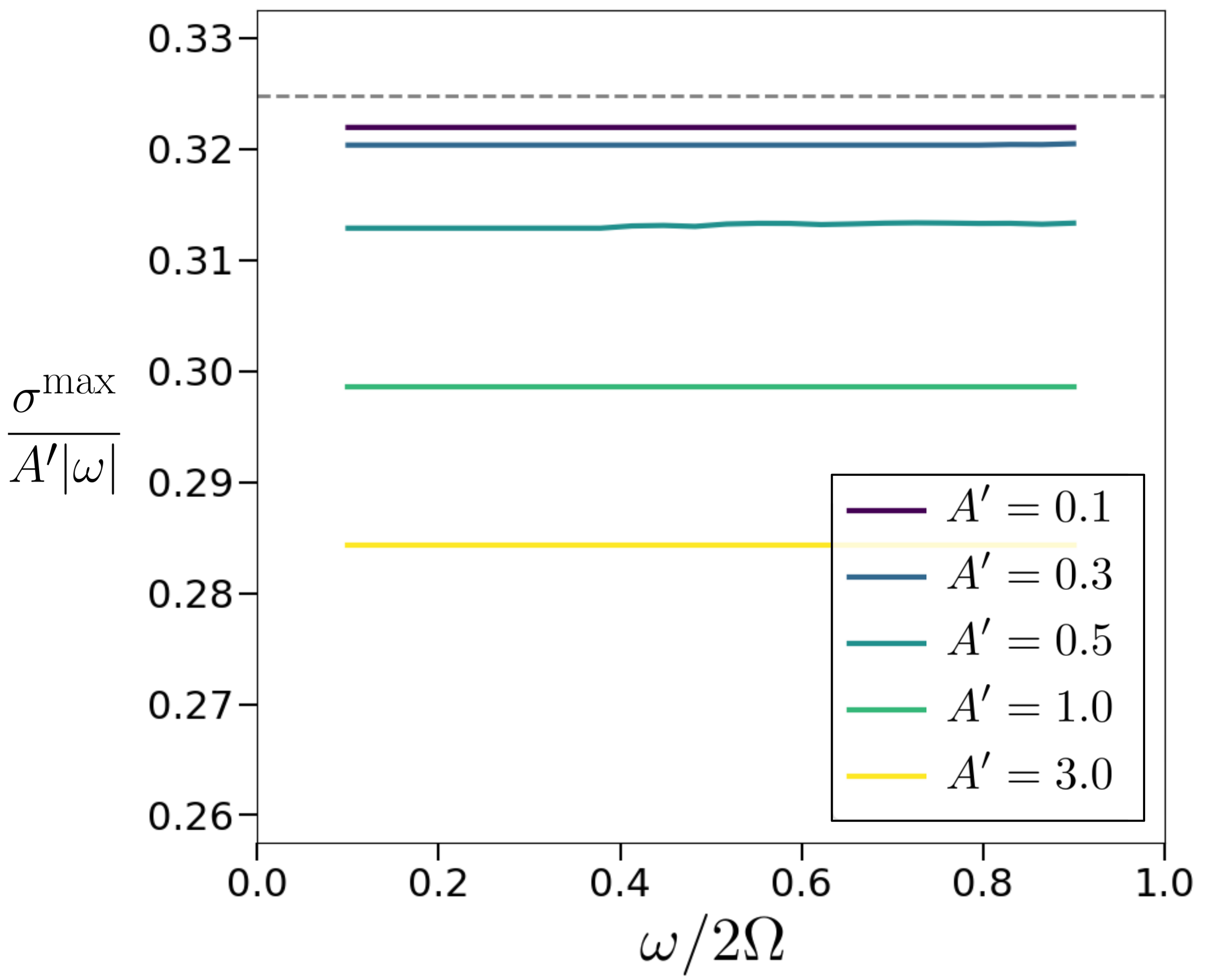}
\caption{\label{fig:sigma_vs_omega}The maximum normalized growth rate $\sigma^{\max}/A'|\omega|$ versus the primary wave frequency $\omega/2\Omega$ for a range of amplitudes $0.1\leq A'\leq 3.0$. For each $\omega$ and $A'$, the $\sigma^{\max}$ is determined by finding the fastest growth rate among perturbations with wave vector components in the range $-16\leq\alpha\leq16$, $0\leq\beta\leq16$, and $\gamma=0$ using the numerical Floquet solver. The magnitude of $\sigma^{\max}/A'|\omega|$ is nearly independent of $\omega$ and is weakly dependent on $A'$. At low $A'\ll1$, the growth rate based on Floquet theory agrees with the maximum growth rate of the PSI, $\sigma_{\rm PSI}/A'|\omega|=3\sqrt{3}/16$, marked by the dashed, gray line.}
\end{figure}

An interesting feature of the Floquet solutions is that the maximum normalized growth rate $\sigma^{\max}/A'|\omega|$ across all $\alpha,\beta$ is nearly independent of both $A'$ and $\omega/2\Omega$, even for $A'\gtrsim 1$. It robustly has a magnitude of $\sim 0.3$. A broader scan of the parameter space $(A',\omega/2\Omega)$ confirms this feature. Figure~\ref{fig:sigma_vs_omega} shows the maximum normalized growth rate across a range of primary wave amplitudes $0.1\leq A'\leq3$ and frequencies $0.1\leq\omega/2\Omega\leq0.9$. At low wave amplitudes, its nearly constant value matches the maximum PSI growth rate $\sigma_{\rm PSI}/A'|\omega|=3\sqrt{3}/16\approx 0.325$ \citep[][see also Appendix \ref{ap:PSI}]{Mora2021}. At high wave amplitudes, the maximum normalized growth rate deviates remarkably little from the PSI prediction (derived for $A'\ll1$), decreasing only slightly to $\approx 0.28$ for wave amplitudes of $A'\sim3$. This is in agreement with the analysis of \cite{AM2024} where the growth rate of short-wavelength perturbations was also found to begin deviating from the PSI prediction for $A'\gtrsim1$. For even larger $A'\gtrsim 3$, the instability growth rate exceeds the primary wave frequency $\sigma^{\max}>|\omega|$, implying that propagating IWs with such large amplitudes would be difficult to observe in any application. 

\section{Simulations of Plane Inertial Waves}
\label{sec:DNS}
Investigating the nonlinear saturation of the instabilities of a propagating IW and the subsequent turbulent evolution of the flow requires using DNS. We carry out DNS in both the comoving and the rotating frames, leveraging the numerical advantages of each frame. Simulations in the comoving frame enable a cross-validation between the DNS and the Floquet solvers. The discrete horizontal wave vector components in the comoving frame of the DNS correspond exactly to the $\alpha,\beta$ used in the Floquet analysis, allowing a direct verification of the Floquet theory. Simulations in the rotating frame are better suited to follow the nonlinear evolution of the flow. The rotating frame has the rotation vector exactly aligned with the numerical grid, allowing geostrophic modes with $k_z=0$ to be exactly captured. Our focus is to examine how the energy of the primary wave is partitioned into the geostrophic and the non-geostrophic components of the turbulent flow as the wave breaks down.

The Navier-Stokes equation for the velocity field is solved as an initial value problem using the Dedalus code \citep{burns2020dedalus}. The domain is a cube of side length $L=16\pi$ with periodic boundary conditions, and the flow is represented as a truncated sum of  Fourier modes $\bs{u}=\sum_{\bs{k}}\bs{u}_{\bs{k}}(t)\exp(\mathrm{i}\bs{k}\cdot \bs{x})$. Time stepping is carried out with a third-order, implicit-explicit Runge-Kutta method \citep{ascher1997implicit} where linear terms are treated implicitly, non-linear terms are treated explicitly, and the timestep $dt$ is set as a small fraction of the shortest growth timescale, $dt\ll1/\sigma^{\max}$. The nonlinear terms are evaluated in real space with the standard $2/3$ dealiasing rule. All runs use a resolution of $N=256^3$ Fourier modes. 

For simulations in the comoving frame, the time-independent base flow $\bs{U}'(\phi)$ is initialized following Equation~(\ref{eq:BaseFlow}), which fits $L/2\pi=8$ wavelengths along the vertical $\hat{\bs{\phi}}$ direction. The wave frequency, $\omega/2\Omega=s\cos\theta$, is controlled by the tilt of the rotation vector $\bs{\Omega}'=-|\tan\theta|\hat{\bs{x}}+(\cos\theta/|\cos\theta|)\hat{\bs{\phi}}$. For simulations in the rotating frame, the base flow $\bs{U}(\bs{x},t=0)$ is initialized following Equation~(\ref{eq:U_lab}).  This results in an IW with a time-dependent flow field $\bs{U}(\bs{x},t)$ propagating inside the domain. The wave frequency is controlled by the choice of wave vector $\bs k=(k_x,0,k_z)$ in the $x$--$z$ plane, where only certain frequencies $\omega/2\Omega=sk_z/k$ (here $k=\sqrt{k_x^2+k_z^2}$) can be chosen based on the discrete Fourier grid. We choose wavenumbers such that $kL/2\pi\sim 8$ to achieve a similar scale separation between the wavelength and the box scale as in the comoving frame. To seed unstable perturbations of the base flow, the velocity field in both setups is additionally initialized with a Gaussian random noise vector field whose solenoidal components are projected away. The Ekman number is chosen to damp unstable modes at the largest wavenumbers in the box $k_{\max}=\pi N/L$ by setting $\Ek=C A'/k_{\max}^2$, where $C\sim1$ is an adjustable ``safety" factor used to control the level of damping. Lastly, the helicity of all primary waves studied below is $s=1$. 

\subsection{Verification of Linear Theory with DNS in the Comoving Frame}
We first cross-validate the DNS in the comoving frame and the Floquet solver by comparing the growth rates obtained independently by the two methods. The growth rates in the DNS are measured by fitting the exponential growth of the energy in the discrete horizontal Fourier modes, which  correspond to the horizontal wave vectors ($\alpha,\beta$) in the Floquet analysis after they have been rescaled by a multiplicative factor of $8$, equal to the ratio of the box scale to the wavelength of the primary wave. The energy $E_{\alpha,\beta}(t)$ at each $(\alpha,\beta)$ is computed by summing over all vertical wavenumbers $k_z$, that is $E_{\alpha,\beta}(t)=0.5\sum_{k_z} |\bs{u}'_{\bs{k}}(\alpha,\beta,k_z,t)|^2$. The growth of $E_{\alpha,\beta}(t)$ is dominated by the fastest growing perturbation with the given $(\alpha,\beta)$ and should closely match the value of $2\sigma^{\max}(\alpha,\beta)$ predicted by Floquet theory. The fitting is done over a time duration $\Delta t\approx 5/\sigma^{\max}$ at the end of the simulation when the instantaneous growth rates of individual modes become steady and approach their maximum values. This avoids the initial transient when slower growing modes can significantly contribute and adversely affect the comparison with theory.

\begin{figure}
    \centering
    \includegraphics[width=\linewidth]{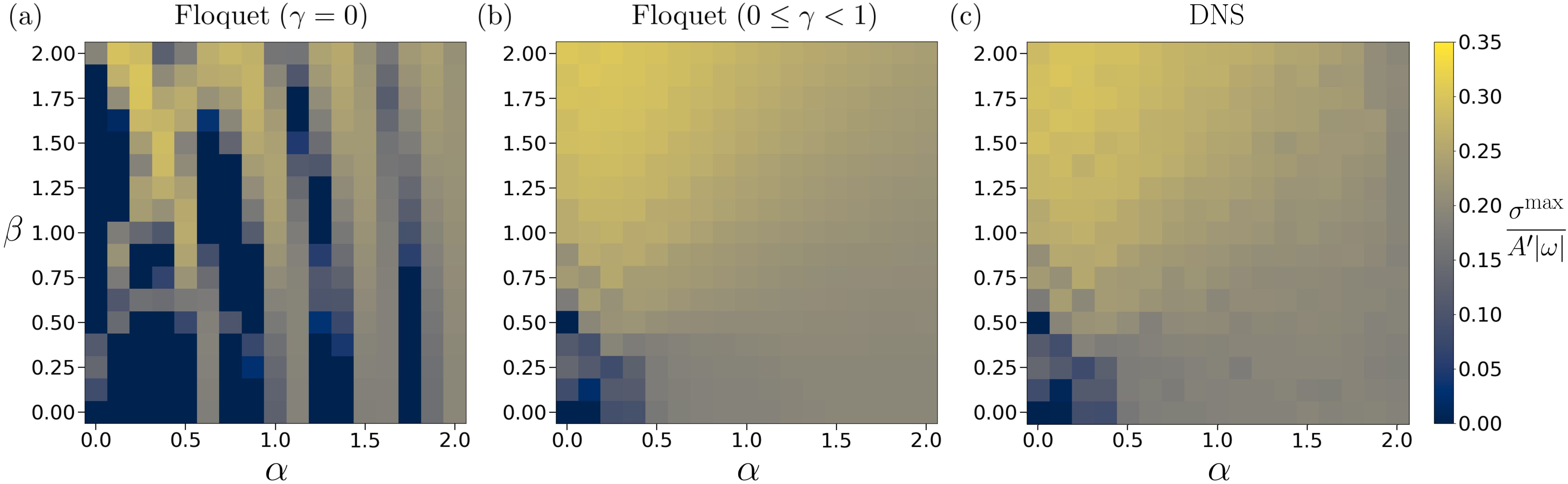}
    \caption{ The DNS in the comoving frame is cross-validated against the Floquet theory for a primary wave with amplitude $A'=0.3$ and frequency $\omega/2\Omega=0.71$ (angle $\theta=45^\circ$). Panels (a) and (b) show the Floquet prediction for $\sigma^{\max}$, where the fastest growing perturbation at each $\alpha,\beta$ is found for fixed $\gamma=0$ in panel (a) and for  all possible $\gamma$ allowed on the discrete Fourier grid of the DNS in panel (b). Panel (c) shows the $\sigma^{\max}$  measured at each $\alpha,\beta$ in the DNS. 
    }
    \label{fig:DNS_Floquet_compare}
\end{figure}

The comparison of $\sigma^{\max}$ in the $\alpha-\beta$ plane is shown in Figure~\ref{fig:DNS_Floquet_compare} for a primary IW with $A'=0.3$, and $\omega/2\Omega=0.71$ (angle $\theta=45^\circ$). The figure shows only the zoomed-in region $0\leq\alpha,\beta\leq2$ for clearer visual inspection. Although the measurement of the growth rates in the DNS can be somewhat noisy (in particular for the slower growing modes), the DNS result (panel c) and the Floquet predictions (panel b) are in good agreement. Note that all available $0\leq\gamma<1$ can be excited and the Floquet solver in panel (b) searches over $\gamma=j/8$ with all possible integers $0\leq j<8$ that exist on the numerical Fourier grid. By contrast, Figures~\ref{fig:omega_scan}, ~\ref{fig:A_scan} and panel (a) of Figure~\ref{fig:DNS_Floquet_compare} only show the maximum growth rate among perturbations with $\gamma=0$. The combined instability ribbons for varying $\gamma$ almost completely cover the $\alpha-\beta$ plane, resulting in a non-zero growth rate at nearly every horizontal wave vector for the $A'$ adopted. The exception is only for low horizontal wavenumbers $\sqrt{\alpha^2+\beta^2}\lesssim 0.5$ (wavelengths exceeding that of the primary IW).

\begin{figure}
\includegraphics[width=\linewidth]{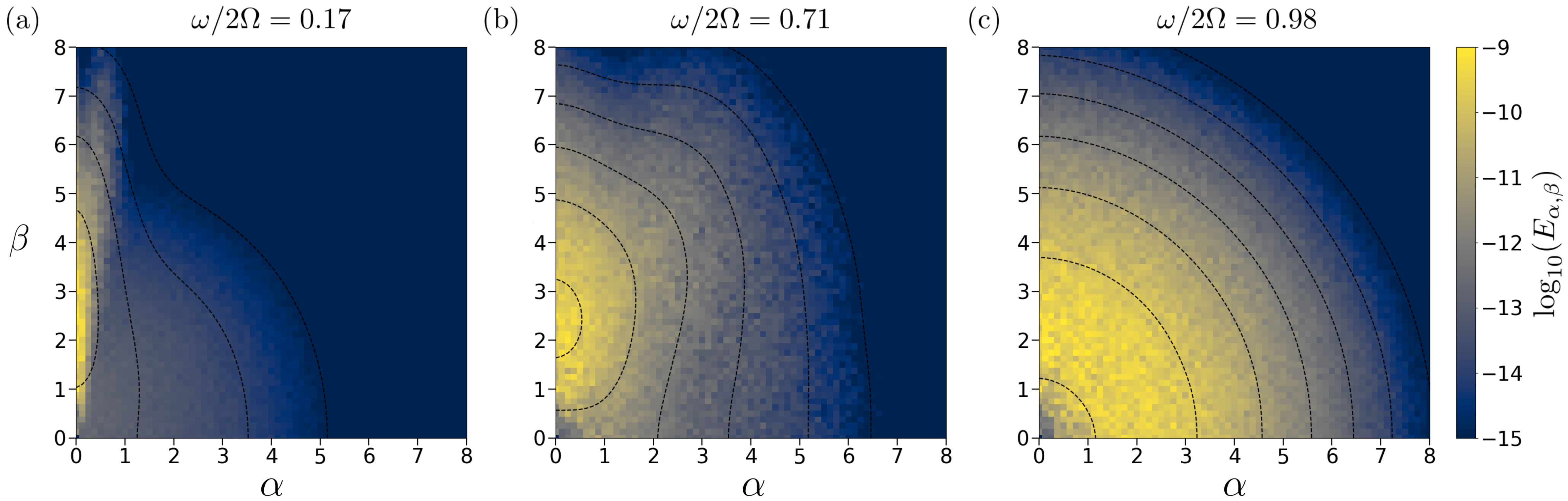}
\caption{\label{fig:Comoving_energyspectra} 2D energy spectra $E_{\alpha,\beta}$ during the linear phase of the DNS in the comoving frame for three primary waves with different frequencies $\omega/2\Omega$, but the same amplitude $A'=0.3$. The fastest growing perturbations gain the most energy, and their distribution in the $\alpha-\beta$ plane depends on the wave frequency in a similar manner as the Floquet predictions in Figure~\ref{fig:omega_scan}. The contours of constant energy are shown as the black dashed lines at integer values of $\log_{10}(E_{\alpha,\beta})=i$ in the range $-15\leq i\leq-9$.
}
\end{figure}

The Floquet theory prediction for how the primary wave frequency affects the wave vector orientation of the most unstable perturbations can also be confirmed by examining the energy spectra $E_{\alpha,\beta}$ for the three fiducial frequencies $\omega/2\Omega=0.17,0.71,0.98$, as shown in Figure~\ref{fig:Comoving_energyspectra}. The energy spectra are computed at similar times during the linear phase of each simulation and the $\alpha,\beta$ with the largest energies correspond to the perturbations that have grown the fastest from their initial values of $E_{\alpha,\beta}\sim 10^{-16}$. One can see that the energy spectra in Figure~\ref{fig:Comoving_energyspectra} qualitatively have the same features and dependence on $\omega/2\Omega$ as the Floquet predictions in Figure~\ref{fig:omega_scan}. The wave vector orientations of unstable perturbations shift from being preferentially aligned with the $\hat{\bs{y}}'$ direction for lower frequency waves to being more isotropically distributed for higher frequency waves. The energy in horizontal wave vectors with wavenumber larger than $\sqrt{\alpha^2+\beta^2}\sim 8$ is small because the viscous damping at these large wavenumbers becomes significant for the Ekman number $\Ek\approx 10^{-3}$ needed to resolve these simulations.

\begin{figure}
    \centering
    \includegraphics[width=\linewidth]{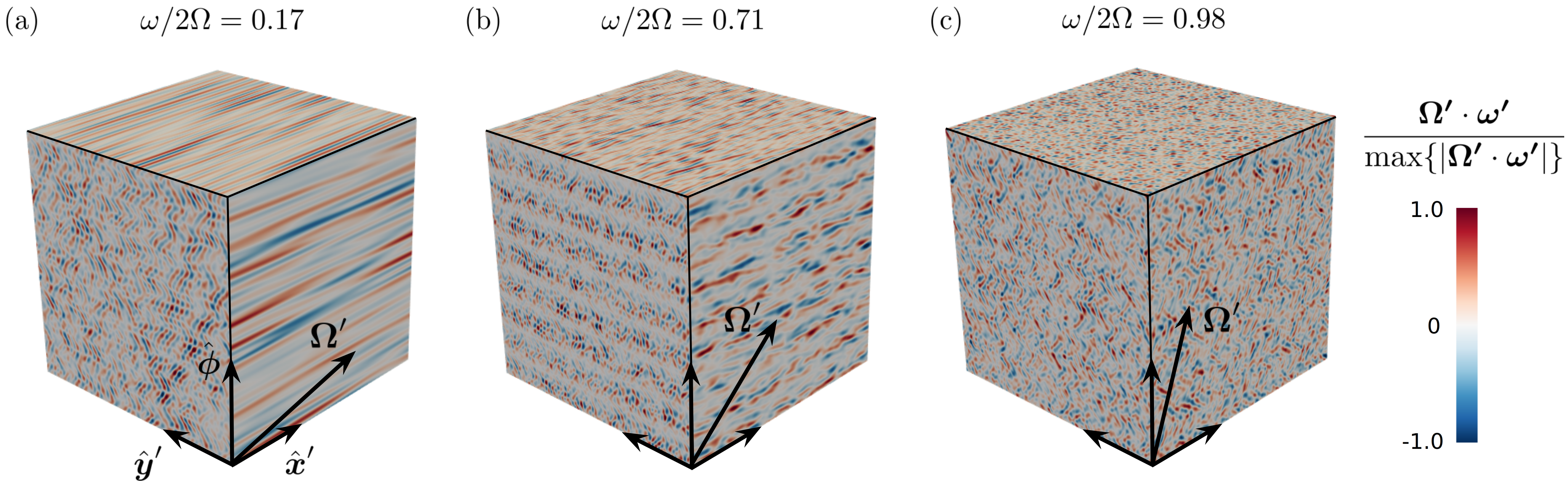}
    \caption{Visualization of vorticity of the perturbations $\bs \omega'=\nabla\times \bs u'$ projected onto the rotation vector $\bs \Omega'$ 
    in the same three DNS in the comoving frame as shown in Figure~\ref{fig:Comoving_energyspectra}.
    The primary IWs have a common amplitude $A'=0.3$, but different frequencies. 
    The vorticity $\bs{\Omega}'\cdot \bs{\omega}'$ in each panel is normalized by its maximum value for visual clarity.   }
    \label{fig:Comoving_3dvis}
\end{figure}

The anisotropy of the wave vector orientation of the most unstable perturbations can also be confirmed in real-space visualizations of the flow. Figure~\ref{fig:Comoving_3dvis} shows the vorticity of the perturbations along the rotation vector $\bs{\Omega}'\cdot \bs{\omega}'$ (where $\bs{\omega}'=\nabla\times\bs{u}'$) for the three fiducial simulations. One can see that the perturbations disturbing the base flow of the low frequency wave $\omega/2\Omega=0.17$ have small structures along the $\hat{\bs{y}}'$ direction, with extended structures that are nearly invariant in the $\hat{\bs{x}}'$ direction. Perturbations of the intermediate frequency wave $\omega/2\Omega=0.71$  also  have  variations primarily in the $\hat{\bs{y}}'$ direction, but also have significant structure in the $\hat{\bs{x}}'$ direction. By contrast, there is no obvious anisotropy in the perturbations of the high frequency wave with $\omega/2\Omega=0.98$.

\subsection{Nonlinear Evolution in the Rotating Frame}

As an IW propagates, unstable perturbations grow at the expense of the wave's energy,  eventually disrupting the flow of the IW and draining its energy. In a rotating fluid, the energy can be converted into two qualitatively different channels. One channel is a forward cascade on dynamical timescales $\sim1/A'|\omega|$ wherein nonlinear interactions progressively transfer energy to smaller scales, ultimately converting the energy into heat through viscous dissipation. The other channel is conversion of the energy into geostrophic modes, which are 2D flows invariant along the rotation axis, with $\bs{\Omega}\cdot\nabla \bs{u}=0$. This conversion in the triply periodic model must occur through near-resonant triadic or resonant quartic interactions, since exact triadic resonances cannot transfer energy from IWs to geostrophic modes due to Greenspan's theorem \citep[e.g.][]{SW1999,Greenspan1969}. The geostrophic motions then inverse cascade through mergers of 2D vortices and decay on long viscous timescales due to their lack of variation in the vertical direction and large horizontal scales \citep[e.g.][]{Godeferd2015}. Below, we explore the evolution of the flow and the partition of the initial wave energy between the geostrophic flow and the forward cascade after the primary IW has broken down.

\begin{figure}
    \centering
    \includegraphics[width=\linewidth]{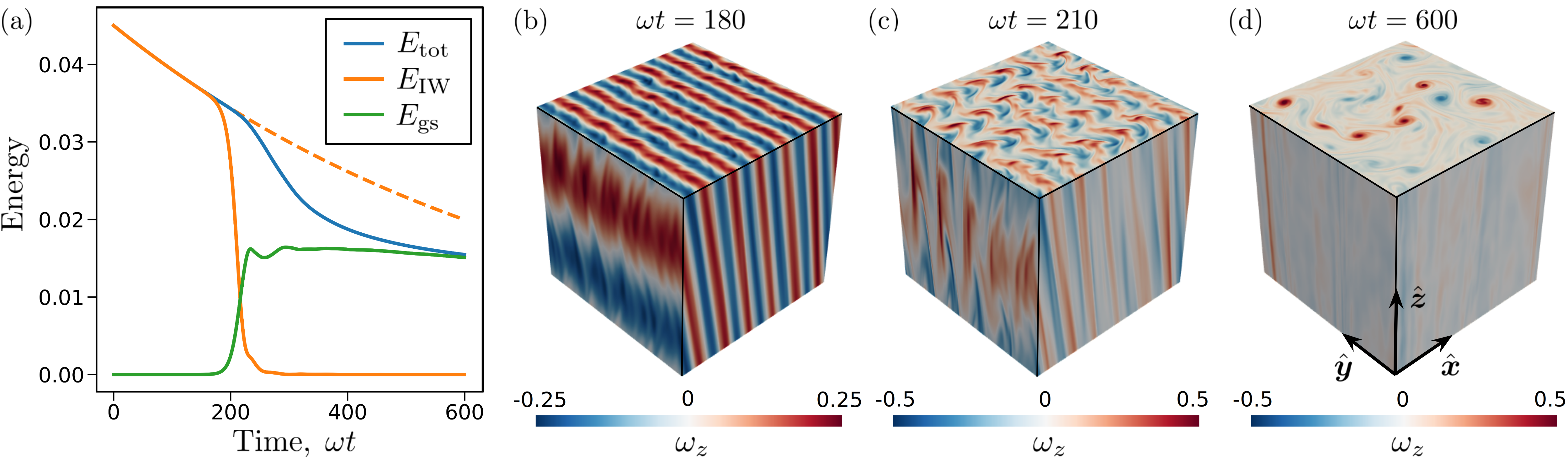}
    \caption{Simulation in the rotating frame of a propagating IW with amplitude $A'=0.3$, frequency $\omega/2\Omega=0.17$ (angle $\theta=80^{\circ}$), and Ekman number $\Ek= 10^{-3}$. Panel (a) shows the temporal evolution of the total energy $E_{\rm tot}$ of the flow field and the contributions from the energy in the primary wave $E_{\rm IW}$ and in the geostrophic flow $E_{\rm gs}$. The dashed line shows $E_{\rm IW}(t)$ if the primary wave decays only by viscous damping $(\propto\exp(-2\Ek k_{\mathrm{IW}}^2 t))$. Panels (b-d) show the vertical vorticity $\omega_z=(\nabla\times\bs{u})\cdot \hat{\bs z}$ (the component aligned with $\bs \Omega=\Omega \hat{\bs z}$) at three times representing the linear, turbulent, and decay phases of the flow evolution. 
    }
    \label{fig:Lab_slices}
\end{figure}

Simulations of propagating IWs are carried out through the nonlinear phase of their evolution in the rotating frame. A visualization of a representative simulation is shown in Figure~\ref{fig:Lab_slices} for a primary IW with $A'=0.3$ and $\omega/2\Omega=0.17$. The flow evolution is split into three distinct phases. First, the linear phase occurs at early times $\omega t\lesssim 180$  when the unstable perturbations grow exponentially (not shown). The viscous damping of the primary wave during this early period is seen as the slow, exponential decay of the IW energy $E_{\rm IW}=0.5|\boldsymbol{u}_{\boldsymbol{k},\rm IW}|^2$ in panel (a). As the unstable perturbations develop significant amplitudes (panel (b) at $\omega t=180$), the flow enters the second, highly turbulent phase $180\lesssim\omega t\lesssim220$ when the unstable modes saturate. This phase is characterized by drainage of the primary wave energy, accompanied by rapid growth of energy in the geostrophic modes $E_{\rm gs}=\sum_{\boldsymbol{k}\;(k_z=0)}|u_{\boldsymbol{k}}|^2$. The flow during this phase has both small-scale features characteristic of a forward cascade and vertically stretched features characteristic of geostrophic motions (panel (c) at $\omega t=210$). The flow then transitions into a third, decay phase where $E_{\rm gs}$ remains nearly constant while the energy in the non-geostrophic flow $E_{\rm tot}-E_{\rm gs}$ (the separation between the blue and green curves in panel (a)) decays through a forward cascade on longer timescales. This phase becomes dominated by the dynamics of merging geostrophic vortices once $E_{\rm gs}\approx E_{\rm tot}$ (panel (d) at $\omega t=600$). Eventually, the vortices will reach the box scale and decay on slow, viscous timescales. Similar behavior is observed in simulations of the elliptical and precessional instabilities \citep[e.g.][]{BL2013,B2016,Pizzi2022,Nils2023}, which continually excite IWs, unlike the decaying case here.

\begin{figure}
    \centering
    \includegraphics[width=0.95\linewidth]{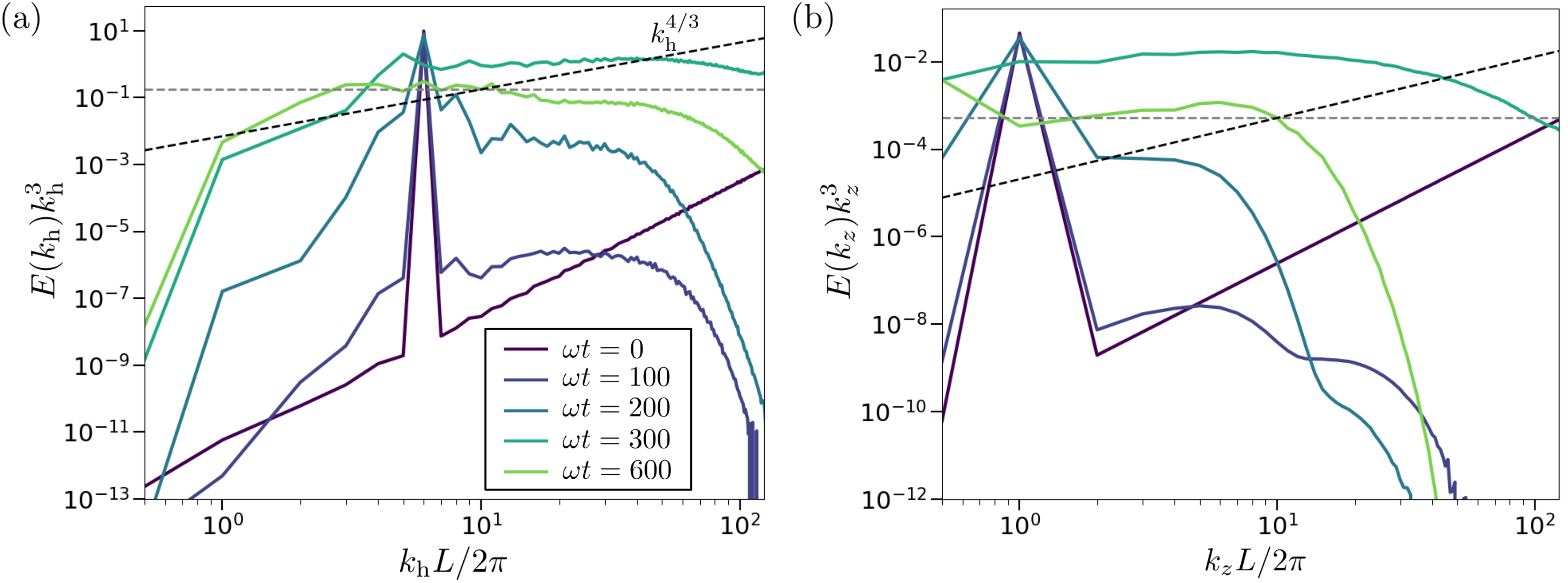}
    \caption{ Energy spectra of the velocity field at representative times (color-coded) during the evolution of a propagating IW. The simulation is the same as the one shown in Figure~\ref{fig:Lab_slices}. Panel (a) shows the horizontal energy spectra $E(k_{\rm h})$ compensated by $k_{\rm h}^{3}$ and panel (b) shows the vertical energy spectra $E(k_{z})$ compensated by $k_z^3$. The compensated horizontal energy spectrum at intermediate scales is nearly flat at late times when the flow is nearly geostrophic.
    }
    \label{fig:Lab_Espectra}
\end{figure}

The three evolutionary phases can also be examined in the 1D horizontal and vertical energy spectra, $E(k_{\rm h})$ and $E(k_z)$, shown in Figure~\ref{fig:Lab_Espectra}, where $k_{\rm h}=\sqrt{k_x^2+k_y^2}$. The initial horizontal and vertical spectra at $ t=0$ have peaks at $k_{\rm h}L/2\pi\sim 6$ and $k_{z}L/2\pi\sim 1$, respectively, corresponding to the primary wave. They also contain power law backgrounds $E(k_{\mathrm{h}})\propto k_{\mathrm{h}}$ and $E(k_{z})\propto k_z^0$ corresponding to the solenoidal Gaussian noise velocity field. At early times $\omega t\lesssim 180$, the energy at the peaks remains unchanged while the energy at other wave vectors is determined by the growth rates of the unstable perturbations (the largest wavenumbers are viscously damped). As the unstable perturbations saturate and the wave breaks down, the horizontal spectrum begins to flatten out during the most turbulent part of the flow's evolution. However, it remains far from a Kolmogorov scaling $E(k_{\rm h})\propto k_{\rm h}^{-5/3}$. At later times, the horizontal energy spectrum approaches the $E(k_{\rm h})\propto k_{\rm h}^{-3}$ scaling expected for geostrophic turbulence \citep[e.g.][]{SW1999}, which appears as a flat compensated spectrum at intermediate scales. The geostrophic-nature of the flow is evidenced in the vertical spectrum by the accumulation of energy in the $k_z=0$ modes at late times (the bin at $k_z L/2\pi=0.5$ in panel (b) holds the energy of modes with $k_z L/2\pi\leq0.5$).

\begin{figure}
    \centering
    \includegraphics[width=0.4\linewidth]{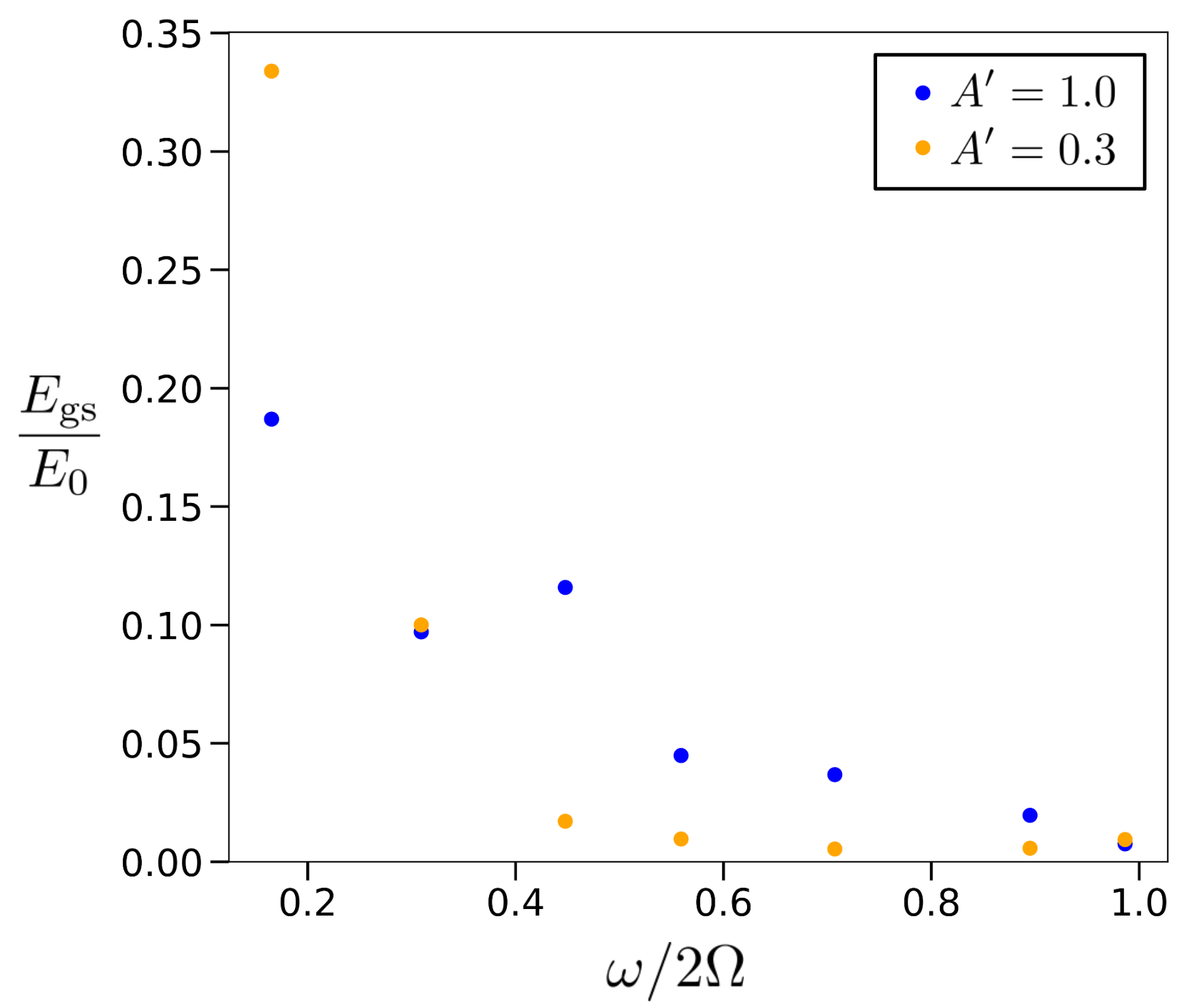}
    \caption{ 
    The fraction $E_{\rm gs}/E_0$ of the initial primary IW energy $E_0$ that is converted into energy of the geostrophic flow $E_{\rm gs}$ in a set of simulations varying the primary IW frequency and amplitude. The measurement of $E_{\rm gs}/E_0$ is made in every simulation long after the primary IW has broken down when $E_{\rm gs}$ contains at least $95\%$ of the residual energy in the flow.
    }
    \label{fig:Lab_GSratio}
\end{figure}

To examine how the nonlinear saturation depends on the properties of the primary IW, we run a set of simulations varying the frequency in a range $0.17\leq\omega/2\Omega\leq0.98$ for two amplitudes $A'=0.3,1.0$. The resulting fraction $E_{\rm gs}/E_0$ of the initial wave energy $E_0$ trapped in the geostrophic component of the flow after wave breakdown is compiled in Figure~\ref{fig:Lab_GSratio}. The measurement of $E_{\rm gs}/E_0$ is done at the end of every simulation when $E_{\rm gs}$ contains at least $95\%$ of the total energy remaining in the flow. One can see that lower frequency waves ($|\omega|\lesssim\Omega$) tend to have a larger portion of their energy transferred into the geostrophic flow, upwards of $E_{\rm gs}/E_0\gtrsim10\%$. On the other hand, higher frequency waves ($|\omega|\gtrsim\Omega$) lose most of their energy to the forward cascade, with only $E_{\rm gs}/E_0\lesssim 5\%$ in the geostrophic flow. This trend may be due to the higher efficiency of nearly-horizontal near-resonance triads $(\omega_p+\omega_q+\omega_r={\cal O}(A'))$ with $\omega_p/2\Omega={\cal O}(A')$, $\omega_q/2\Omega={\cal O}(A')$, $\omega_r=0$, in transferring energy into the geostrophic mode \citep{SW1999}, where the low frequency primary wave participates as one of the nearly-geostrophic modes. We caution that part of the non-monotonic trend of $E_{\rm gs}/E_0$ with $\omega$ for $A'=1.0$ at lower frequencies (where the wave vector of the primary wave is highly inclined) may be due to box-size effects, as the number of (near) resonant triads available for interactions may be limited by the discreteness of the numerical Fourier grid. 

The conversion fraction $E_{\rm gs}/E_0$ also depends on the wave amplitude. Lower amplitude waves tend to have a smaller conversion fraction, at least for higher frequency waves with $\omega/2\Omega\gtrsim0.3$. This trend is consistent with some of the proposed mechanisms that can transfer energy into geostrophic modes with a rate $\sigma_{\rm gs}\sim A'^2|\omega|$ when $A'\lesssim1$, which include near-resonant triadic \citep{newell1969rossby,SW1999,smith2005near,di2016quantifying,LeReun2020} and resonant four-mode interactions \citep{newell1969rossby,Kerswell1999,SW1999,brunet2020shortcut}.  These mechanisms have been used to explain the steady-state fraction of energy in the geostrophic component of forced rotating turbulence, and we expect that these same mechanisms operate during the breakdown of a single IW. The scaling $\sigma_{\rm gs}/\sigma_{\rm PSI}\propto A'$ suggests that at sufficiently low wave amplitudes, the pumping of geostrophic modes will be too slow to meaningfully operate within the timescale for wave breakdown $\sim1/\sigma_{\rm PSI}$.  However, we leave exploration of the saturation of lower amplitude waves with $A'\lesssim 0.1$ for future work due to the increased computational cost of simulating flows with low Ekman numbers, including the longer run time required for the instabilities to grow and saturate. 

Lastly, we note that the conversion fraction $E_{\rm gs}/E_0$ may also depend on the choice of initial perturbations, including their amplitude and distribution in wave vector space. Larger amplitude  noise reduces the amount that the primary wave decays due to viscosity because unstable modes saturate earlier. It also allows slower growing modes to participate in the saturation phase, in contrast to the case of extremely small noise amplitude where mainly the fastest growing modes contribute. Despite these effects, we have found that the final value of $E_{\rm gs}/E_0$ is weakly altered for large noise amplitudes, changing by at most a few percent for initial noise energy up to $(E_{\rm tot}-E_{0})/E_{\rm tot}\sim0.01$. We leave exploring different noise spectra, or possibly seeding single unstable eigenmodes, for future work.

\section{Conclusion}\label{Conclusion}

Our analysis combined a comprehensive numerical Floquet method (complemented with an analytical asymptotic analysis) with direct numerical simulations to investigate the linear stability and nonlinear saturation of propagating plane IWs. We found that the fastest-growing unstable perturbations have a non-dimensional growth rate $\sigma^{\max}/(A\,k)\approx 0.3 $ that is nearly independent of the wave amplitude, where $A$ is the velocity amplitude and $k$ is the wavenumber. This rate closely matches the growth rate of the PSI at low amplitudes $A\ll|\omega|/k$  (where $\sigma_{\mathrm{PSI}}/(A\,k)\approx 0.325 $) and decreases only slightly by around $\sim15\%$ at larger amplitudes $A\sim|\omega|/k$. We also found that the most unstable wavenumbers are much larger than the wavenumber of the primary IW at low wave amplitudes, as expected for the PSI, but become comparable with the primary's wavenumber at high amplitudes. Furthermore, the orientation of the wave vector of the most unstable modes strongly depends on the primary IW frequency. For low-frequency waves $|\omega|\ll2\Omega$, the wave vectors are aligned orthogonal to the plane of the rotation axis and the primary IW wavevector, while for high-frequency waves $|\omega|\lesssim2\Omega$, their distribution becomes nearly isotropic. This suggests that a single IW is generally disrupted by perturbations that isotropize the flow in the plane perpendicular to $\bs{\Omega}$ as the wave breaks down, in close agreement with the findings of \citep{Mora2021} and \citep{AM2024}.

We then used DNS to study the breakdown of propagating IWs in a triply periodic domain. Our simulations first cross-validated the Floquet theory and then explored the nonlinear evolution of the flow. When the primary IW breaks down, its energy can be transferred into two complementary channels. It can either be dissipated through a forward cascade on dynamical timescales or accumulated in slowly evolving geostrophic modes. The relative efficiency of these two channels depends on the interactions between the primary IW and the saturating unstable perturbations. We found that lower-frequency waves generally convert a larger fraction of initial wave energy into geostrophic motions (e.g., a fraction of more than $10\%$ for IWs with $|\omega|\lesssim\Omega$ and high amplitudes with $A\sim |\omega|/k$). In contrast, higher-frequency waves lose most of their energy through the forward cascade. The conversion fraction to geostrophic energy also decreases as the wave amplitude decreases, with the exception of the lowest-frequency waves. 

The turbulence resulting from the breakdown of a single IW may share aspects of forced rotating turbulence, which pumps geostrophic modes when the Rossby number is above a threshold value. As geostrophic modes cannot gain energy from triadic resonance interactions in this triply periodic model, energy accumulation in geostrophic modes likely occurs through near-resonant triadic interactions \citep{newell1969rossby,SW1999,smith2005near,di2016quantifying,LeReun2020} or higher-order resonant interactions \citep{newell1969rossby,Kerswell1999,SW1999,brunet2020shortcut}, which are less efficient at lower Rossby numbers. We suggest that studying the breakdown of IWs could be a cleaner numerical experiment for understanding these processes.  Discerning the threshold Rossby number (wave amplitude) and the mechanisms for pumping geostrophic modes during the breakdown of a single IW requires future work. 

The linear stability problem and simulations can also be extended to include more realistic backgrounds, such as stable stratification, differential rotation, or magnetic fields.  Stratification is particularly important in oceanic environments where IWs are excited predominantly by wind stresses and are a major source of mixing in the upper oceans \citep{alford2016near}, impacting oceanic biogeochemistry and the climate \citep{jochum2013impact}. In astrophysical contexts, IWs excited by planet-star tidal interactions can propagate in the convection zones of their host stars (or inside giant planets themselves) where differential rotation and/or large-scale magnetic fields modify wave propagation and instability \citep[e.g.][]{OL2007,AB2022,AB2023,AB2025}. In close orbits or when excited in resonance, IWs are expected to achieve large amplitudes, and understanding their general propagation, stability, and dissipation is key for predicting tidal evolution.

We note that in realistic physical systems, either in experimental or natural fluids, IWs are typically excited in beams of finite width. The linear instability analysis in our triply periodic model is applicable when unstable daughter waves, whose group velocities would be misaligned with the primary wave beam, have a growth timescale that is shorter than the time it takes them to travel out of the beam. It is well known that finite-size effects in IGW beams restrict the wave vector space of unstable modes that can cause wave breakdown \citep{koudella2006instability,bourget2013experimental,bourget2014finite,dauxois2018instabilities,grayson2022long} and thus affect how far an IGW beam propagates before disrupting and depositing its energy and momentum. Our results for the linear stability of IWs can be similarly extended to understand the propagation and transport properties of finite-amplitude IW beams.

\begin{acknowledgments}
AA was supported by a Leverhulme Trust Early Career Fellowship and by STFC grant ST/S000275/1. AJB was supported by STFC Grant No. ST/S000275/1, ST/W000873/1 and UKRI1179. We would like to thank the Isaac Newton Institute for Mathematical Sciences, Cambridge, for support and hospitality during the programs DYT2 and “Anti-diffusive dynamics: from sub-cellular to astrophysical scales”, supported by EPSRC Grant No. EP/R014604/1. We would also like to thank the Kavli Foundation and the Max Plank Institute for Solar System Research at the University of Goettingen who supported the ``Kavli Summer Program in Astrophysics 2021: Fluid Dynamics of the Sun and Stars," during which this project was initiated.
\end{acknowledgments}

\section*{Data Availability}
The data underlying this article will be shared on reasonable request to the corresponding author.

\appendix

\section{Energy conservation equation for the Floquet system}
\label{ap:EnergyConsrv}
The kinetic energy evolution of the perturbations $K'(\phi)=\left(|\hat u|^2+|\hat v|^2+|\hat w|^2\right)/2$ is obtained by using the Floquet ansatz (Eq.~(\ref{eq:Floquet_2})) in the momentum equation (Eqs.~\ref{eq:NSx}, \ref{eq:NSy}, and \ref{eq:NSz})  to express it in terms of periodic $\hat f$ variables, and then multiplying by the complex conjugates of the velocity components $\hat u^*$, $\hat v^*$, and $\hat w^*$, respectively. This gives
\begin{align}
    2\mathrm{Re}[\sigma' ]K'+W'\frac{\mathrm{d}K'}{\mathrm{d}\phi}=&-\mathrm{Re}[\hat u^* \hat w]\frac{\mathrm{d}U'}{\mathrm{d}\phi}-\mathrm{Re}[\hat v^*\hat w]\frac{\mathrm{d}V'}{\mathrm{d}\phi}-\mathrm{Im}\left[\alpha \hat p^*\hat u+\beta \hat p^* \hat v+\gamma\hat p^*\hat w\right]-\mathrm{Re}\left[\frac{\mathrm{d}\hat p^*}{\mathrm{d}\phi}\hat w\right]\nonumber\\
    &+\Ek\,\mathrm{Re}\left[\hat u^*L_2\hat u+\hat v^*L_2 \hat v+\hat w^* L_2 \hat w \right], \label{eq:bal}
\end{align}
This equation is similar to the energy transfer equation for propagating IGWs derived in \citet[Eq. (20)]{LR1996}, except without buoyancy fluxes. The first two terms on the right-hand side of Eq. (\ref{eq:bal}) are shear production or extraction terms, which can act to transfer energy between the primary wave and the perturbations. By contrast, the advective term on the left-hand side ($\propto W'$) and the two pressure terms on the right-hand side are conservative and vanish when integrated over one wavelength. Carrying out the integration gives the relation between the growth rate and the various source and sink terms:
\begin{align}
2\mathrm{Re}[\sigma' ]=-\left\langle\mathrm{Re}[\hat u^* \hat w]\frac{\mathrm{d}U'}{\mathrm{d}\phi}\right\rangle_\phi-\left\langle\mathrm{Re}[\hat v^*\hat w]\frac{\mathrm{d}V'}{\mathrm{d}\phi}\right\rangle_\phi+\Ek\,\left\langle\mathrm{Re}\left[\hat u^*L_2\hat u+\hat v^*L_2 \hat v+\hat w^* L_2 \hat w \right]\right\rangle_\phi, \label{eq:balvol}
\end{align}
where $\langle\cdot\rangle_\phi=\int^{2\pi}_0\cdot\,\mathrm{d}\phi\,/(2\pi \left\langle K'\right\rangle_\phi)$. The growth rate can be positive when the shear terms are positive (extracting energy from the primary wave), and the viscous dissipation term is relatively small. 

\section{Theory for small  amplitude waves: TRI and PSI}
\label{ap:PSI}

The stability of a propagating IW can be analyzed analytically for small wave amplitudes ($A'\ll1$) using the method of multiple scales \citep{BenderOrszag} (for a similar approach in other problems, see ~\citep{MK2002,OL2013,BO2014,CuiLatter2022}). The method takes advantage of the large timescale separation between the short oscillation period of the primary wave and the long growth timescale of infinitesimal perturbations, which turn out to also be IWs. As our focus is to obtain the maximum growth rates of perturbations, viscous effects are neglected in the analysis below. 

We work in the comoving frame where the perturbed velocity field $\bs u$ satisfies Equation~(\ref{eq:NS_2b}), reproduced below:
\begin{align}
    \label{eq:ap_NS_1}
    &(\partial_t+W\partial_\phi)\bs{ u}+\bs {\Omega}'\times \bs{ u}+\bm\nabla p=-\bs{U}_p\cdot\bm\nabla\bs {u}-\bs{ u}\cdot\bm\nabla\bs {U}_p.
\end{align}
The prime notation for the coordinates and the velocity field in the comoving frame has been dropped, and the primary wave flow has been redefined as
\begin{align}
    \bs{U}=\bs{U}_p+W\hat{\bm\phi},\quad \bs{U}_p=
    (U,V,0).
\end{align}

The multiple-scale analysis  formally introduces a ``slow" timescale $\tau=\epsilon t$ that is treated as if it is independent of the ``fast" timescale $t$ throughout the calculation, until replacement in the final solution. The ordering parameter $\epsilon\ll 1$ quantifies the smallness of the wave amplitude $A'$. We seek a regular asymptotic expansion of the form 
\begin{align}
\bs{u}&=\bs{u}_0(x,y,\phi,t,\tau)+\epsilon \bs{u}_1(x,y,\phi,t,\tau)+\dots.
\end{align}
Substituting the above ansatz into Equation~(\ref{eq:ap_NS_1}), taking $A'={\cal O}(\epsilon )$, and using $\partial_t \to \partial_t + \epsilon \partial_\tau$, the zeroth $\mathcal{O}(1)$ and first-order $\mathcal{O}(\epsilon)$ terms in the momentum equation are
\begin{align}
\label{eq:epsilon_order1a}
\epsilon^0:\quad (\partial_t+W\partial_\phi)\bs{ u}_0+\bs {\Omega}'\times \bs{ u}_0+\bm\nabla p_0=& 0, \\
\label{eq:epsilon_order2a}
\epsilon^1:\quad (\partial_t+W\partial_\phi)\bs{ u}_1+\bs {\Omega}'\times \bs{ u}_1+\bm\nabla p_1=&-\bs{\mathcal{F}},\quad \bs{\mathcal{F}}=\partial_\tau \bs{u}_0+\bs{U}_p\cdot\bm\nabla\bs {u}_0+\bs{ u}_0\cdot\bm\nabla\bs {U}_p.
\end{align}

The dynamical equation at each order of $\epsilon$ can be conveniently rewritten in the form of a Poincar\'e-like equation, where the pressure is eliminated using the incompressibility constraint. Taking the curl twice and applying the derivative operator $-(\partial_t+W\partial_\phi)$ at each order gives their Poincar\'e-like form:
\begin{align}
\label{eq:epsilon_order1}
\epsilon^0:\quad \mathcal{L}\bs{u}_0&=\bm 0, \\
\label{eq:epsilon_order2}
\epsilon^1:\quad \mathcal{L}\bs{u}_1&=\left(\partial_t+W\partial_\phi\right)\bm\nabla\times\bm\nabla\times \bs{\mathcal{F}}+\bs {\Omega}'\cdot\bm\nabla\left(\bm\nabla\times \bs{\mathcal{F}}\right)\equiv\bs{G}(\bs{u}_0),
\end{align}
where $\bs{G}$ is a source term and $\mathcal{L}$ defines the linear operator 
\begin{align}
    \label{eq:ap_Poinc_1}
    \mathcal{L}\equiv\left(\partial_t+W\partial_\phi\right)^2\bm\nabla^2+\left(\bs {\Omega}'\cdot\bm\nabla\right)^2.
\end{align}

The general solution to the zeroth order equation (\ref{eq:epsilon_order1}) is a linear superposition of plane-wave solutions oscillating on fast timescales with amplitudes evolving on slow timescales. This can be written in the form 
\begin{align}
&\bs{u}_0=\sum_X X(\tau)\bs u_X E_X(\bs x,t)+\mathrm{c.c.},\quad E_X\equiv\mathrm{e}^{\mathrm{i}(-\omega_X t + \bs{k}_X\cdot\bm x)},
\label{eq:planewave}
\end{align}
where $\mathrm{c.c.}$ denotes the complex conjugate and $X$ denotes a given wave. Each wave satisfies the dispersion relation
\begin{align}
&\left(\omega_X-Wk_{X,\phi}\right)^2-\frac{(\bs{\Omega}'\cdot \bs{k}_X)^2}{k_X^2}=0,\quad k_X^2=k_{X,x}^2+k_{X,y}^2+k_{X,\phi}^2.
\label{eq:dispRelA}
\end{align}
This is simply the dispersion relation for IWs in the comoving frame, where the frequency is Doppler-shifted by the factor $W\hat{\bm\phi}\cdot \bs{k}_X$. The corresponding eigenvector (arbitrary amplitude chosen for convenience) is given by
\begin{align}
\bs{u}_X=\left(1,
\frac{-\mathrm{i}\tilde\omega_X k_{X,y}k_{X,\phi}-k_{X,x}(\bs{\Omega}'\cdot\bs{k}_X)}{-\mathrm{i}\tilde\omega_X k_{X,x}k_{X,\phi}+k_{X,y}(\bs{\Omega}'\cdot\bs{k}_X)},
\frac{-\mathrm{i}\tilde\omega_X k_{X,\phi}k_{X,y}+k_{X,x}(\bs{\Omega}'\cdot\bs{k}_X)}{-\mathrm{i}\tilde\omega_X k_{X,x}k_{X,y}-k_{X,\phi}(\bs{\Omega}'\cdot\bs{k}_X)}\right),\quad \tilde\omega_X\equiv\omega_X-Wk_{X,\phi}.
\end{align}
Determining the amplitude evolution on slow timescales (e.g. finding $X(\tau)$) requires substitution of $\bs{u}_0$ into $\bs{G}$ in the first-order equation (\ref{eq:epsilon_order2}) and examining the solvability conditions for the first-order solution to maintain $\epsilon\bs{u}_1=O(\epsilon)$ at all times. This requires $\bs{G}(\bs{u}_0)$ to not contain terms proportional to homogeneous solutions of the linear operator $\mathcal{L}\bs{u}_1=0$, which are terms proportional to $E_X$. 

The solvability condition can lead to unstable solutions (whose amplitudes $X(\tau)$ grow exponentially) in the particular case of a triadic resonance, which we focus on below. This occurs when $\bs{u}_0$ is the superposition of two waves ($\omega_a,\bs{k}_a$) and ($\omega_b,\bs{k}_b$) (with flow $\bs{u}_0 = a(\tau) \bs{u}_a E_a +b(\tau) \bs{u}_b E_b + c.c$) that satisfy temporal and spatial resonance with the primary wave $(\omega_p,\bs{k}_p)$. The resonance conditions are given by
\begin{align}
    \pm\omega_p\pm\omega_a\pm\omega_b=0,\; \pm\bs{k}_p\pm\bs{k}_a\pm\bs{k}_b=\bs{0}.
\end{align}
In the comoving reference frame where $\omega_p=0$ and $\bs{k}_p=(0,0,1)$, these become 
\begin{align}
    \label{eq:res_cond_2}
     \omega_a&=\omega_b, \; k_{a,x}=k_{b,x},\; k_{a,y}=k_{b,y},\;k_{a,\phi}=k_{b,\phi}-1,
\end{align}
where we have chosen plus in front of $a$ and minus in front of $b$ throughout  without loss of generality (note that $\omega_{a,b}$ and $\bs{k}_{a,b}$ can be of either sign).

\begin{figure}
    \centering
    \includegraphics[width=0.95\linewidth]{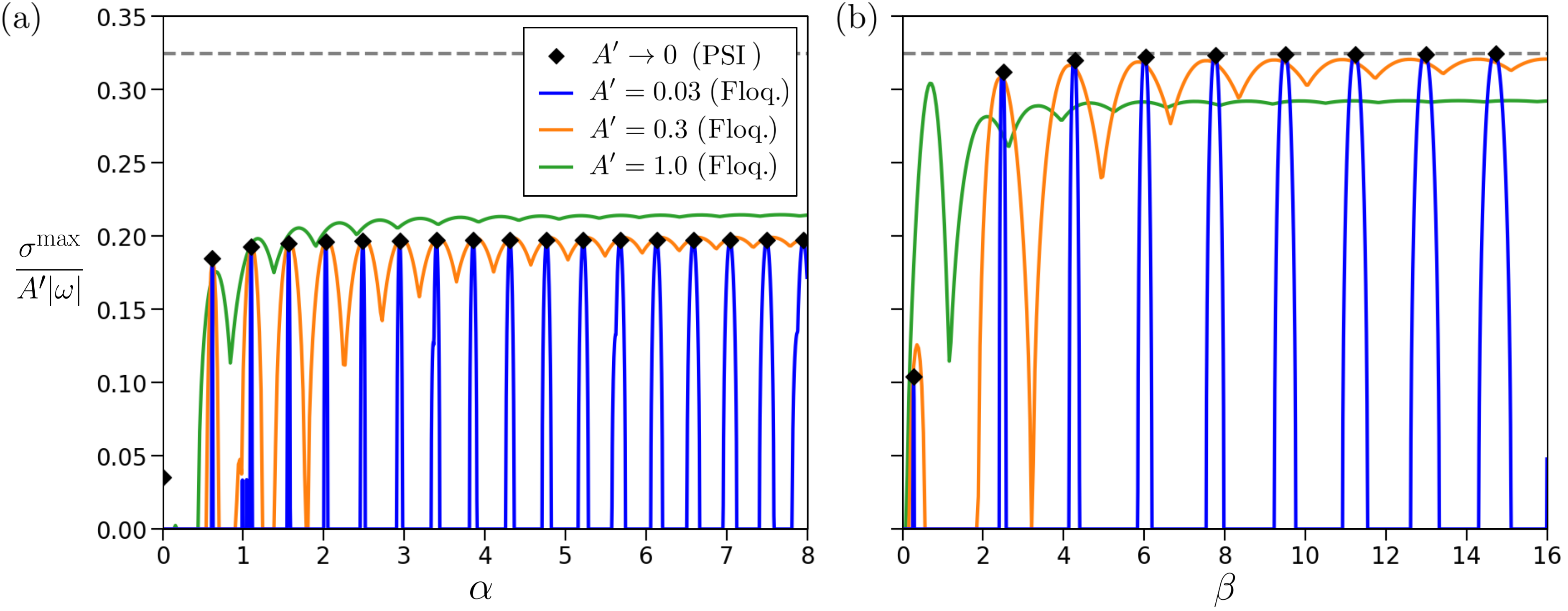}
    \caption{The fastest growing unstable perturbations according to the PSI and the Floquet analysis for various amplitudes $A'$ of a primary IW with fixed frequency $\omega/2\Omega=0.71$ ($\theta=45^\circ$). Panel (a) shows $\sigma^{\max}$ versus $\alpha$ at fixed $\beta=0$ and $\gamma=0$. Panel (b) shows $\sigma^{\max}$ versus $\beta$ at fixed $\alpha=0$ and $\gamma=0$. The PSI and Floquet theories agree at low wave amplitudes. However, they deviate at larger amplitudes when couplings of the $n\textendash$th mode with modes beyond $n\pm1$ become significant in the Floquet theory, which invalidates the three-wave assumption made in the PSI analysis. The Floquet solutions for $A'=0.3$ in panels (a) and (b) are identical to a horizontal slice at $\beta=0$ and a vertical slice at $\alpha=0$, respectively, through panel (b) in Figure~\ref{fig:omega_scan}. The fastest growth rate of the PSI, $\sigma_{\rm PSI}\approx 0.325,$ is shown as the dashed, gray line, which closely matches the growth rates of the perturbations with $\alpha=0$ in panel (b) (see Section~\ref{sec:NumericalFloq_1} for discussion).
    }
    \label{fig:PSI}
\end{figure}

Computing $\bs{G}(\bs{u}_0)$ requires the intermediate step of computing the vector $\bs{\mathcal{F}}$, which we find is
\begin{align} 
\nonumber
    \bs{\mathcal{F}}(\bs{u}_0)=& (\partial_\tau a) \bs{u}_a E_a + \mathrm{i}a\left[(\bs{\hat U}_p\cdot\bs{k}_a)\bs{u}_aE_p+(\hat{\bs{U}}^*_p\cdot\bs{k}_a)\bs{u}_aE_p^*+(\bs{u}_a\cdot\bs{k}_p)\hat{\bs{U}}_pE_{p}-(\bs{u}_a\cdot\bs{k}_p)\hat{\bs{U}}^*_pE_p^*\right]E_a \\ &
    +(\partial_\tau b) \bs{u}_bE_b +  \mathrm{i}b\left[(\bs{\hat U}_p\cdot\bs{k}_b)\bs{u}_bE_p+(\hat{\bs{U}}^*_p\cdot\bs{k}_b)\bs{u}_bE_p^*+(\bs{u}_b\cdot\bs{k}_p)\hat{\bs{U}}_pE_p-(\bs{u}_b\cdot\bs{k}_p)\hat{\bs{U}}^*_p E_p^*\right]E_b + \mathrm{c.c.},
\end{align}
where we have defined $\hat{\bs{U}}_p=(A'/2)(1,-s\mathrm{i},0)$ so that $\bs{U}_p=\hat{\bs{U}}_p E_p+\hat{\bs{U}}^*_p E_p^*$. This can then be used to find the terms $\nabla\times\bs{\mathcal{F}}$ and $\nabla\times \nabla\times\bs{\mathcal{F}}$ needed to compute $\bs{G}$. Note that $\bs{\mathcal{F}}$ (and its derivatives) only contains terms $E_a, E_b, E_aE_p, E_aE_p^*, E_bE_p, E_bE_p^*$ which are related by the products $E_pE_a=E_b$ and $E_p^*E_b=E_a$, since $E_XE_X^*=1$. 

Gathering terms, one can write
\begin{align}
    \bs{G}(\bs{u}_0) = \bs{g}_a E_a+\bs{g}_bE_b +\dots,
\end{align}
where $\bs{g}_a$ and $\bs{g}_b$ contain only constants, $a(\tau)$ and $b(\tau)$, and their slow time derivatives. The three dots correspond to terms proportional to $E_p^*E_a$ or $E_pE_b$ that do not satisfy a resonance condition. We find
\begin{align}
\bs{g}_a =& -2\mathrm{i}\tilde{\omega}_a k_a^2\partial_\tau a\bs{u}_a+ b\left\{-\mathrm{i}\bs{\Omega}'\cdot \bs{k}_a\left[
(\hat{\bs{U}}^*_p\cdot\bs{k}_b) \bs{k}_a\times \bs{u}_b - (\bs{u}_b\cdot\bs{k}_p) \bs{k}_a\times \hat{\bs{U}}^*_p\right]\right. \nonumber\\
&
\left.+\tilde{\omega}_a\left[
(\hat{\bs{U}}_p^*\cdot \bs{k}_b)\left( (\bs{k}_p\cdot\bs{u}_b)\bs{k}_a+k_a^2\bs{u}_b\right) + (\bs{u}_b\cdot\bs{k}_p)\left( (\bs{k}_b\cdot\hat{\bs{U}}^*_p)\bs{k}_a-k_a^2\hat{\bs{U}}_p^*\right)\right]
\right\},
\label{gA} 
\end{align}
where we have used the resonance condition $\bs{k}_b-\bs{k}_p=\bs{k}_a$. The result is similar for $\bs{g}_b$, but with the replacements $a\rightarrow b$, $\bs{\hat U}_p\leftrightarrow \bs{\hat U}_p^*$, and $\bs{k}_p\rightarrow -\bs{k}_p$. 

The solvability condition requires each of the Cartesian components of the vectors $\bs{g}_a$ and $\bs{g}_b$ to vanish (so that $\bs{G}(\bs{u}_0)$ contains no terms proportional to $E_a$ or $E_b$), giving
\begin{align}
    \partial_\tau a - \mathrm{i}C_{ab} b=0, \;\partial_\tau b + \mathrm{i}C_{ba} a=0,\label{eq:odetau}
\end{align}
where $C_{ab}$ and $C_{ba}$ are determined from the coefficients in $\bs{g}_a$ and $\bs{g}_b$ (i.e., Equation~(\ref{gA})). An exponential solution $a,b\propto \exp(\sigma' \tau)$ has growth rate \footnote{In the cases discussed below and in Fig.~\ref{fig:PSI}, where either $k_{a,x}=\alpha=0$ or $k_{a,y}=\beta=0$, we found that the coefficients $C_{ab}$ and $C_{ba}$ (and hence $\sigma$) are both real. We have also confirmed this numerically for selected values of $\alpha, \beta$ and $\gamma$ when evaluating the growth rate based on the analysis in this appendix.} $\sigma'=\pm\sqrt{C_{ab}C_{ba}}$ such that the dimensional growth rate is $\sigma=\pm\sqrt{C_{ab} C_{ba}}A'|\omega|$, where we have equated $\epsilon$ with the primary wave amplitude. The derivation leading to Eq.~\ref{eq:odetau} is straightforwardly valid as long as the coefficient of $\partial_\tau a$ in Eq.~(\ref{gA}) does not vanish, in which case it must be considered more carefully. This coefficient reduces to $-2\mathrm{i}\tilde\omega_a k_a^2\bm u_a$, which vanishes when $\omega_a=k_{a,\varphi} W$, corresponding to the perturbation comoving with the primary wave, somewhat like a ``critical layer" or ``corotation resonance".

We expect that for small $A'$, the maximum growth rate is achieved for modes that most precisely satisfy the resonance conditions in Eq.~(\ref{eq:res_cond_2}), which is generally easier at large perturbation wavenumbers $\alpha^2+\beta^2+(\gamma+n)^2\gg1$. Here, we connect with the notation in the Floquet analysis by using $\alpha,\beta$ for the horizontal components of the daughter wave's wave vector and $\gamma+n$ for the component along $\hat\phi $ (where $n$ is an integer and $0\leq\gamma<1$). In this limit with $\bs{k}_a\approx \bs{k}_b$, the TRI reduces to the PSI where the daughter waves have half the frequency of the primary wave in the rotating frame. For simplicity, and motivated by the numerical results, we consider modes with $\alpha=0$ and $s=1=k_z/|k_z|$ (so $W=-1$) such that $\bs{k}_a=(0,\beta,\gamma+n)$ and $\bs{k}_b=(0,\beta,\gamma+n+1)$, thus satisfying the resonance condition Eq.~(\ref{eq:res_cond_2}), with $k_a^2=\beta^2+(\gamma+n)^2$ and $k_b^2=\beta^2+(\gamma+n+1)^2$. The eigenvectors ($X=a,b$) simplify to
\begin{align}
\bs{u}_X=\left(1,-\frac{\mathrm{i}\tilde{\omega}_X k_{X,\phi}}{\bs{\Omega}'\cdot\bs{k}_X},\frac{\mathrm{i}\tilde{\omega}_X k_{X,y}}{\bs{\Omega}'\cdot\bs{k}_X}\right)=\left(1,-\mathrm{i}\tilde{\omega}_X ,\pm\frac{i k_{X,y}}{k_{X}}\right).
\end{align}
From the dispersion relation Eq.~(\ref{eq:dispRelA}), the frequencies are given by $\tilde\omega_a^2=(\gamma+n)^2/k_a^2$ and $\tilde\omega_b^2=(\gamma+n+1)^2/k_b^2$ since $\bs{\Omega}'\cdot\bs{k}_a=\gamma+n$ and $\bs{\Omega}'\cdot\bs{k}_b=\gamma+n+1$. To satisfy the resonance conditions, we decide to set $\tilde{\omega}_a=-(\gamma+n)/k_a$ and $\tilde{\omega}_b=(\gamma+n+1)/k_b$. The expressions for $C_{ab}$ and $C_{ba}$, and hence $\sigma'$, can then be obtained as a function of $\beta$ and $\gamma+n$. The large wavenumber limit of the product $C_{ab}C_{ba}$ is taken by fixing the wave vector orientation $r=\beta/(\gamma+n)$ for a constant ratio $r$ and then taking the limit $\beta\to\infty$ (or equivalently, $n\to\infty$). We obtain
\begin{align}
    \sqrt{C_{ab}C_{ba}}\to \frac{1}{4}\left(\frac{r}{1+r^2}+\frac{r}{\sqrt{1+r^2}}\right).
\end{align}
The maximum growth rate occurs for $r\to \pm \sqrt{3}$ and has a dimensional value of
\begin{align}
\sigma_{\mathrm{PSI}} = \frac{3\sqrt{3}}{16}A'|\omega|.
\end{align}
This growth rate for PSI is consistent with the results obtained in \cite{Mora2021,AM2024}. They find $\sigma_{\mathrm{PSI}}=0.25 \sin\alpha_2(1+\cos\alpha_2)A'|\omega|=(3\sqrt{3}/16)A'|\omega|$ with the inclination angle of the wavevector $\alpha_2=\pi/3$ corresponding to $\tan{\alpha_2}=r=\sqrt{3}$.  Hence, our results are in agreement. This multiple-scale analysis is also numerically consistent with the Floquet theory as shown in Figure~\ref{fig:PSI}. Note that this PSI limit does not necessarily lead to the fastest growth rate over all possible modes -- in that we have not proven that it is the optimum growth rate over all $\beta$ and $n$, though we expect this to be the case on physical grounds.

\bibliography{apssamp}

\end{document}